\documentclass[fleqn,usenatbib,useAMS]{mnras}

\usepackage[T1]{fontenc}
\usepackage{ae,aecompl}

\usepackage{graphicx}	% Including figure files
\usepackage{amsmath}	% Advanced maths commands
\usepackage{amssymb}	% Extra maths symbols
\usepackage{xspace}
\newcommand{\mhp}{$M_{\rm h}^{\rm peak}$ }
\newcommand{\mhpp}{$M_{\rm h}^{\rm peak}$}
\newcommand{\msmh}{$M_\ast/M_{\rm h}$ }
\newcommand{\msmhp}{$M_\ast/M_{\rm h}$}
\newcommand{\mhmax}{\ensuremath{M_{\rm{h, max}}}\xspace}
\title[Stellar-to-halo mass relationship in COSMOS]{The COSMOS-UltraVISTA stellar-to-halo mass relationship: new insights on galaxy formation efficiency out to $z\sim5$}

\author[L. Legrand et al.]{
L. Legrand$^{1, 2},$\thanks{E-mail: louis.legrand@ias.u-psud.fr}
H. J. McCracken$^{2}$, I. Davidzon$^{3}$, O. Ilbert$^{4}$, J. Coupon$^{5}$,
\newauthor{N. Aghanim$^{1}$, M. Douspis$^{1}$, P.\ L. Capak$^3$, O. Le F\`evre$^{4}$, B. Milvang-Jensen$^{6,7}$}
\\
\\
$^1$Institut d'Astrophysique Spatiale, CNRS (UMR 8617), Universit\'e Paris-Sud, B\^{a}timent 121, Orsay, France \\
$^2$Institut d'Astrophysique de Paris, Sorbonne Universit\'es, UPMC Univ Paris 6 et CNRS, UMR 7095, 98 bis bd Arago, 75014 Paris\\
$^3$IPAC, Mail Code 314-6, California Institute of Technology, 1200 East California Boulevard, Pasadena, CA 91125, USA\\
$^4$Aix Marseille Univ, CNRS, CNES, LAM, Marseille, France\\
$^5$Department of Astronomy, University of Geneva, Ch. d'Ecogia 16, 1290 Versoix, Switzerland\\
$^6$Cosmic Dawn Center (DAWN), Niels Bohr Institute, University of Copenhagen, Juliane Maries Vej 30, DK-2100 Copenhagen \O; \\
$^7$DTU-Space, Technical University of Denmark, Elektrovej 327,
DK-2800 Kgs.\ Lyngby\\
}

\date{}

\pubyear{2018}

\begin{document}
\label{firstpage}
\pagerange{\pageref{firstpage}--\pageref{lastpage}}
\maketitle
\date{\fbox{\sc Draft Version: \today}}
\begin{abstract}
Using precise galaxy stellar mass function measurements in the COSMOS field we determine the stellar-to-halo mass relationship (SHMR) using a parametric abundance matching technique. The unique combination of size and highly complete stellar mass estimates in COSMOS allows us to determine the SHMR over a wide range of halo masses from $z\sim0.2$ to $z\sim5$. At $z\sim 0.2$ the ratio of stellar-to-halo mass content peaks at a characteristic halo mass $M_{\rm h} =10^{12} M_\odot$ and declines at higher and lower halo masses. This characteristic halo mass increases with redshift reaching $M_{\rm h} =10^{12.5} M_\odot$ at $z\sim2.3$ and remaining flat up to $z=4$. We considered the principal sources of uncertainty in our stellar mass measurements and also the variation in halo mass estimates in the literature. We show that our results are robust to these sources of uncertainty and explore likely explanation for differences between our results and those published in the literature. The steady increase in characteristic halo mass with redshift points to a scenario where cold gas inflows become progressively more important in driving star-formation at high redshifts but larger samples of massive galaxies are needed to rigorously test this hypothesis. 
\end{abstract}


%%%%%%%%%%%%%%%%%%%%%%%%%%%%%%%%%%%%%%%%%%%%%%%%%%

%%%%%%%%%%%%%%%%% BODY OF PAPER %%%%%%%%%%%%%%%%%%

\section{Introduction}
\label{sec:introduction}
Galaxy formation is a remarkably inefficient process \citep[e.g.,][]{Silk1977,Persic&Salucci1992,Dayal+2018}. This can be seen quantitatively if one compares the dark matter halo mass function and the galaxy stellar mass function: both in low-- and high--mass regimes they differ by several orders of magnitude \citep[see e.g.][]{Cole+2001,Yang+2003,Eke+2006,Behroozi+2010,Moster+2010}.

Understanding how the stellar mass content ($M_*$) of a galaxy relates to the mass of its dark matter halo ($M_{\rm h}$) is, in fact, an alternative way of considering the problem of galaxy formation. In the local Universe, there is a ``characteristic halo mass" (\mhpp) at which the \msmh ratio is maximised. A natural interpretation is that \mhp corresponds to the halo mass at which star formation, integrated over the entire assembly history of the galaxy, has been the most efficient \citep{Silk+2013}. We consider ``galaxy formation efficiency" as the global process of forming stars in dark matter haloes, from the accretion of gas to the actual transformation of baryons into stars. At lower and higher halo masses, the \msmh ratio decreases rapidly, presumably as a consequence of physical processes that suppress star formation in these haloes. Various mechanisms have been proposed in order to explain this inefficiency: for example, supernovae and stellar winds in low-mass haloes and active galactic nuclei (AGN) feedback processes in more massive objects (see \citealp{Silk+2012} for a detailed review). 

Although such comparisons between mass functions are phenomenological in nature \citep{Mutch+2013} they provide useful constraints to theoretical models of galaxy formation in particular when the comparison spans a large redshift range. The advent of highly complete, mass-selected galaxy surveys \citep[see, e.g.,][]{Ilbert+2013} and accurate predictions for the halo mass function \citep{Tinker+2008,Watson+2013,Despali+2016} allows us to measure the stellar-to-halo mass relationship (SHMR) of galaxies at different epochs. There are many techniques to accomplish this: e.g., in the ``sub-halo abundance matching"  the number density of galaxies (from observations) and dark matter sub-haloes (from simulations) are matched to derive the SHMR at a given redshift \citep[see, e.g.,][]{Marinoni&Hudson2002,Behroozi+2010,Behroozi+2013_SHMR,Behroozi+2018,Moster+2010,Moster+2013,Moster+2018,Reddick+2013}. This technique can also be implemented by assuming a non-parametric monotonic relation between the luminosity or stellar mass of the observed galaxies and sub-halo masses at the time of their infall onto central haloes  \citep{Conroy+2006}.

Other studies use a ``halo occupation distribution" modeling \citep[HOD, see e.g.][]{Vale+2004,Zheng+2007, Leauthaud+2011,Coupon+2015} where a prescription for how galaxies populate dark matter haloes can be used to simultaneously predict the number density of galaxies and their spatial distribution. In this case, lensing combined with clustering measurements can provide additional constraints on the SHMR.

However, until now, investigations of the SHMR over a large redshift range have mostly relied on heterogeneous datasets each with their own selection functions. Interpreting these results can be challenging since different biases from each survey may introduce artificial trends. In this work, we measure the SHMR and \mhp in ten bins of redshifts between $z=0.2$ and $z=5.5$ in a homogeneous and consistent way using the sub-halo abundance matching technique applied to a single dataset: the COSMOS2015 galaxy catalogue \citep{Laigle+2015}.

COSMOS \citep{Scoville+2007} is a 2\,deg$^2$ field with deep UV-to-IR coverage \citep[see][and references therein]{Laigle+2015}. The wealth of spectroscopic observations \citep{Lilly+2007,LeFevre+2005,LeFevre+2015,Hasinger+2018} means photometric redshifts can be validated even in the traditionally poorly-sampled $1<z<2$ redshift range (see figure 11 of \citeauthor{Laigle+2015} and figure 4 of \citeauthor{Davidzon+2017}). The large area of COSMOS make it ideal to collect robust statistics of distant, massive galaxies. Moreover, exquisite IR photometry means precise stellar mass estimates can be made over a large redshift range \citep[see e.g.][]{Steinhardt+2014,Davidzon+2017}.
Extensive tests have been made to validate the mass completeness and the photometric redshift accuracy in COSMOS \citep{Laigle+2015, Davidzon+2017}.
Far-IR, radio, and X-ray observations are also available to assess the crucial role of AGN \citep{Delvecchio+2017},  and the quenching of distant and massive galaxies \citep{Gozaliasl+2018}.

Previously in COSMOS \citet{Leauthaud+2012} used a combination of parametric abundance matching, galaxy clustering and galaxy-galaxy lensing to derive the SHMR to $z\sim1$; galaxy-galaxy lensing measurements with COSMOS ACS data are not feasible above $z\sim1$. More recently, \citet{Cowley+2018} made a  halo modelling analysis to derive the SHMR in the UltraVISTA ''deep stripes`` region.

The organisation of the paper is as follows. In Section \ref{sec:MassFunctions} we introduce the observed stellar mass function of COSMOS galaxies and discuss the principal uncertainties; we then present the \citet{Despali+2016} dark matter halo mass function we use and our fit using a dark matter simulation to derive the halo mass function for the maximum mass in the history of the haloes. We also present comparisons with other mass functions for consistency checks. In Section \ref{sec:SHMR} we describe our abundance matching technique, its assumptions and principal sources of uncertainties, along with our Monte Carlo  Markov Chain (MCMC) fitting procedure. In Section \ref{sec:Results} we present our results, i.e.\ the SHMR and its redshift evolution up to $z\sim5$. We discuss the physical mechanisms that may explain our observations in Section \ref{sec:Discussion}.

Throughout this paper we use the \emph{Planck} 2015 cosmology \citep{Planck+2016} with $\Omega_{\rm m,0} = 0.307$, $\Omega_{\Lambda, 0} = 0.691$, $\Omega_{\rm b,0} = 0.0486$, $N_{\rm eff}=3.05$, $n_{\rm s}= 0.9667$, $h = H_0/(100 \ \rm km \, s^{-1} \, Mpc^{-1}) = 0.6774$, except if noted otherwise.
Stellar mass scales as $1/h^{2}$ whereas halo mass scales as $1/h$. The notation $\phi$ will denote a mass function. The notation $\ln()$ refers to the natural logarithm and $\log()$ refers to the base 10 logarithm.

\section{Mass functions and their uncertainties}
\label{sec:MassFunctions}

\subsection{Stellar mass functions}
\label{sec:SMF}

The galaxy stellar mass function (SMF) corresponds to the number density per unit comoving volume of galaxies in bins of $M_*$. It is one of the key demographics to understand quantitatively the galaxy formation process as it describes how stellar mass is distributed in galaxies. Traditionally, the SMF has been modelled by a \citet{Schechter+1976} function, although for certain galaxy populations a combination of more than one such function may provide a better fit to observations \citep{Binggeli+1988,Kelvin+2014}. Here, we use SMFs derived by \citet[][hereafter D17]{Davidzon+2017} for galaxies in the UltraVISTA-Ultra deep region of the COSMOS field \citep[see][]{McCracken+12}. The sample was constructed using the photometric catalogue of \citet{Laigle+2015} which contains more than half a million galaxies with photometric redshifts ($z_\mathrm{phot}$) between $z=0.2$ and $z=6$ (178,567 of them in the Ultra deep region).
By restricting the analysis to the high-sensitivity region ($K_\mathrm{s}<24.7$\,mag at 3$\sigma$, $\sim$0.7\,mag deeper than the rest of COSMOS) the effective area turns out to be  $\sim$0.5\,deg$^2$. Nonetheless, this represents a 3$\times$ larger volume than the one probed by other deep extragalactic surveys like the Cosmic Assembly Near-IR Deep Extragalactic Legacy Survey \citep[CANDELS,][]{Grogin+2011,Koekemoer+2011}.

Both $z_\mathrm{phot}$ and $M_*$ are derived by fitting the galaxy spectral energy distribution (SED) with synthetic templates (see D17 for further details).
The unique combination of deep optical (Subaru), near-infrared (VISTA) and mid-infrared (\emph{Spitzer}/IRAC) observations results in a galaxy sample that is $>$90\% complete at $M_*>10^{10} M_{\odot}$ up to $z=4$; for galaxies at $4<z<6$ above that  threshold, the catalogue is $>$70\% complete. More generally, D17 defined a  minimal mass ($M_{*,\rm min}$) as the $75\%$ completeness limit, with a redshift evolution  described as $M_{*,\rm min}(z) = 6.3 \times 10^{7} (1+z)^{2.7}M_{\odot}$. This minimal mass is used as the lower boundary for the SMF. 

\begin{figure}
\includegraphics[width=0.99\columnwidth]{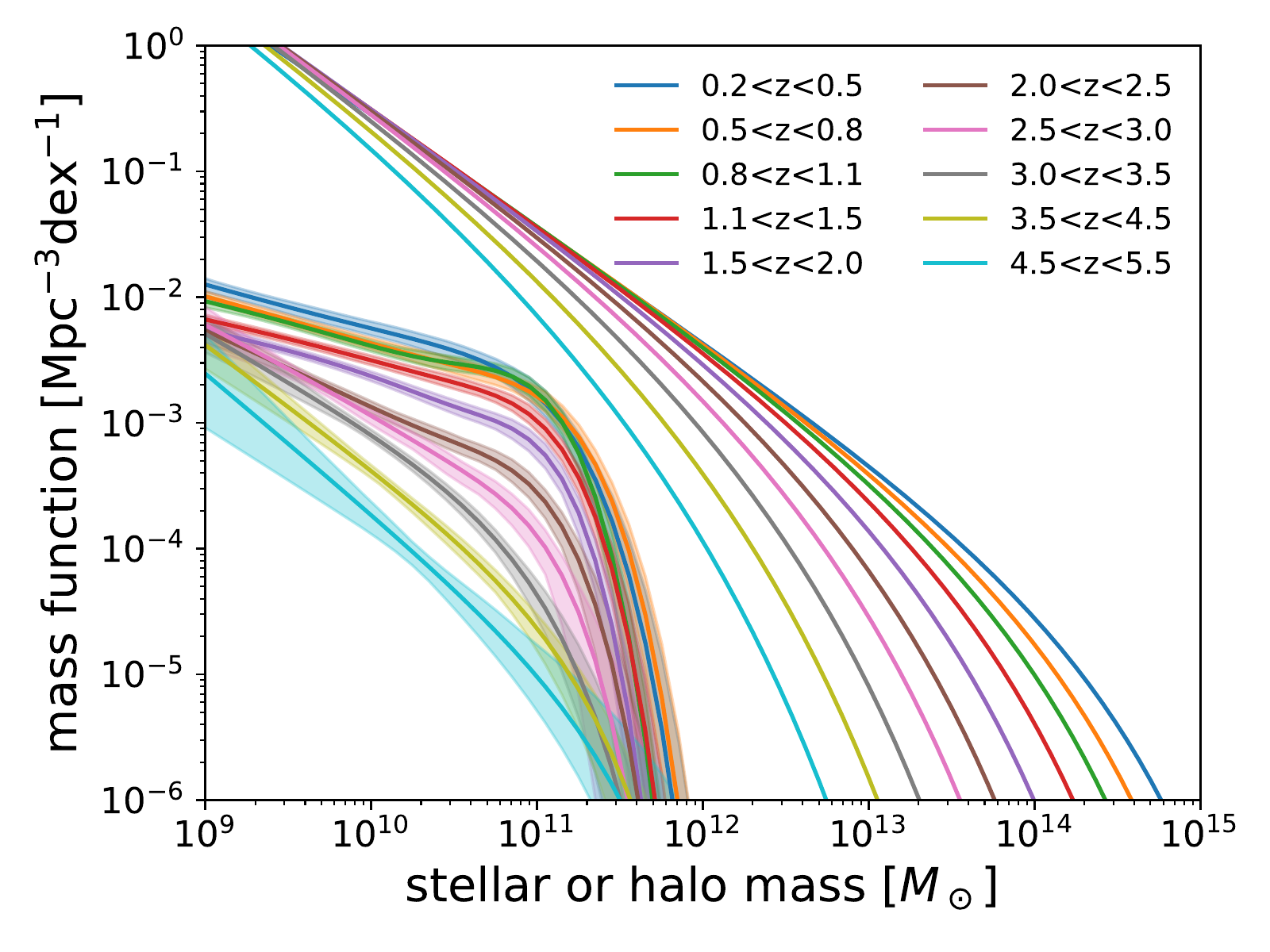}\\
\includegraphics[width=0.99\columnwidth]{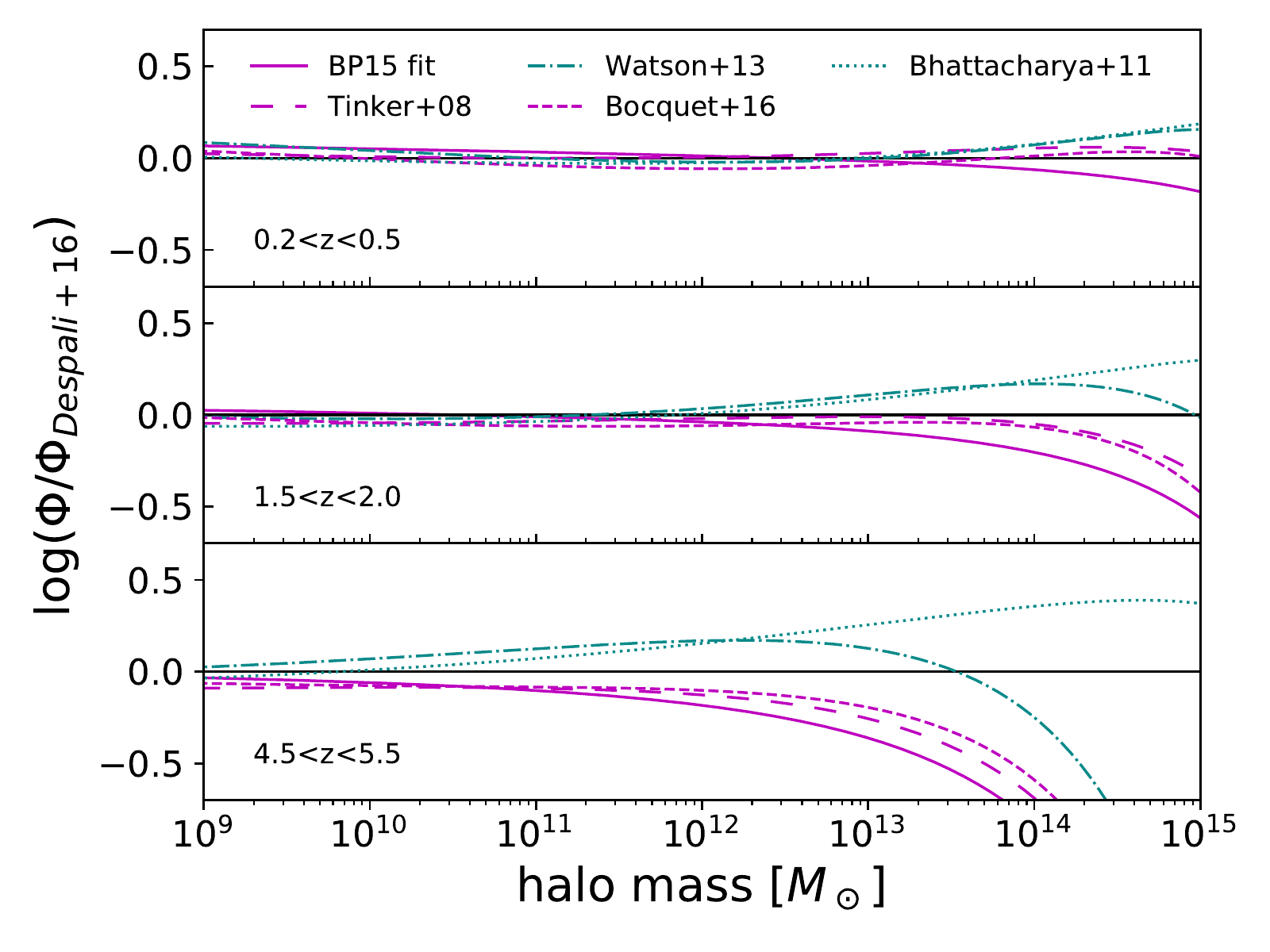}
\caption{\textit{Upper panel:} Our adopted stellar and halo mass functions. For the SMF at a given redshift (see legend) a solid line shows our HMF \citep[][fitted on the Bolsho\"{i}-Planck simulation]{Despali+2016} whilst the solid line and shaded area is the SMF with the associated 1$\sigma$ uncertainty \citep[corresponding to the best fit to $1/V_\mathrm{max}$ points  corrected for Eddington bias,][]{Davidzon+2017}.   \textit{Lower panel:} For three sample redshift bins the relative difference as a function of halo mass between the original \citet{Despali+2016} using the virial overdensity criterion, our Bolsho\"{i}-Planck fit (solid magenta line), and a selection of HMFs from the literature. Magenta lines show numerical simulations in which haloes are defined according to a spherical over-density threshold (solid line:  Bolsho\"{i}-Planck, long-dashed line: \citealt{Tinker+2008}, short dashed: \citealt{Bocquet+2016}). Cyan lines show works that use a friends-of-friends algorithm (dotted line: \citealt{Bhattacharya+2011}, dot-dashed: \citealt{Watson+2013}).}
\label{fig:SMF&HMF}
\end{figure}

D17 estimated the SMF in 10 redshift bins from $z=0.2$ to $z=5.5$ (see Fig.\ \ref{fig:SMF&HMF}) using three independent methods:  the $1/V_{\rm max}$ technique \citep{Schmidt1968}, the step-wise maximum likelihood \citep{Efstathiou+1988} and the maximum likelihood method of \citet{Sandage+1979}. These three estimators provide consistent SMF estimates. However, they are all affected by observational uncertainties ($M_*$ and $z_\mathrm{phot}$ errors) that scatter galaxies from their original mass bin. This systematic effect, known as \citet{Eddington1913} bias, dominates at high masses ($M_*\gtrsim10^{11} M_\odot$) because here galaxy number density declines exponentially; this produces an asymmetric scatter and consequently modifies the SMF profile. Depending on the ``skewness''  and the magnitude dependence of  observational errors the Eddington bias may have a strong impact also at lower masses \citep{Grazian+2015}.

When fitting a \citet{Schechter+1976} function to their $1/V_{\rm max}$ determinations, D17 account for the Eddington bias using the method introduced in  \citet{Ilbert+2013}. Therefore  in our work we use the  Schechter fits of D17 which should be closer to the intrinsic SMF compared to the other estimators. For consistency, we rescale these estimates  to \citet[][P16]{Planck+2016} cosmology.   The fitting function assumed by D17 is a double Schechter (see Eq.\ 4 in D17) at $z<3$ and a single Schechter function (their Eq.\ 3) above that redshift. At low redshifts two SMF components are clearly visible \citep[e.g.][]{Ilbert+2010}, above $z>3$ there is no evidence of this double Schechter profile \citep{Wright+2018}.

The SMF error bars include both systematic and random errors including Poisson noise, cosmic variance \citep[computed using an updated version of the software described in][]{Moster+2011} and the scatter due to errors in the SED fitting. The SMF uncertainties due to SED fitting are derived through Monte Carlo re-extraction of $z_\mathrm{phot}$ and $M_*$ estimates according to the likelihood function of each galaxy. This procedure may be biased if the likelihood  were under- or over-estimated by the SED fitting code \citep[see][]{Dahlen+2013}. However, recent work with simulated photometry suggests that this should not be the case for the code used in D17 (Laigle et al., in prep.).

\subsection{Halo mass functions}

\label{sec:HMF}

Our main reference for the dark matter halo mass function\footnote{HMFs were computed using the \textsc{Colossus} \texttt{python} module \citep{Diemer+2018}.} (HMF) is the work of \citet[][see Fig.\ \ref{fig:SMF&HMF}]{Despali+2016}. They measure the HMF using six $N$-body cosmological simulations with different volumes and resolutions:  all of them have $1024^3$ dark matter particles with masses ranging from $1.94\times10^7$ to $6.35\times10^{11}\,h^{-1}\,M_\odot$ and a corresponding box size from 62.5 to 2000\,$h^{-1}$\,Mpc.
Haloes are identified through  the ``spherical overdensity'' algorithm \citep{Press&Schechter1974}, i.e.\ each halo is a sphere with a matter density equal to the virial overdensity \citep[see][]{Eke1996} at the given redshift  (which is equal to the median $z$ of the observed SMF, see Table \ref{tab:MCMC_fits}). 
The halo mass is defined as the sum of dark matter particles included in such a sphere.

It has been shown \citep[see e.g.][]{Reddick+2013} that for abundance matching applications the stellar mass of galaxies is better correlated to the maximal mass the dark matter haloes have over their history (\mhmax) rather than the actual mass at a given redshift. This is particularly true for sub-haloes which can lose mass due to gravitational stripping by the neighbouring main halo whilst the galaxy inside will keep the same stellar mass. \citet{Reddick+2013} has demonstrated that using this \mhmax better fits to observations such as galaxy clustering for abundance matching.

Our halo mass function for the maximal mass \mhmax are calculated using the Bolsho\"{i}-Planck simulation \citep[][]{Rodriguez-Puebla+2016, Behroozi+2018}. This dark-matter-only simulation has a comoving volume of $250 h^{-1} \rm Mpc$ on a side with $2048^3$ particles with a mass resolution of $1.6 \times 10^8 h^{-1} M_\odot$ and uses \citet{Planck+2016} cosmology. Haloes are identified with the \textsc{Rockstar} halo finder and masses are computed using the virial overdensity criterion of \citet{Bryan+1998}.
\citet{Behroozi+2018} provides halo number densities for several halo mass bins and for 178 snapshots from $z=16$ to $z=0$ for this simulation. We fit the HMF of \citet{Despali+2016} using a modified version of the \textsc{Colossus} code for these data points in the range $0<z<5$ and $10^{11} h^{-1} M_\odot < \mhmax < 10^{15} h^{-1} M_\odot$. The parameters of equation 7 of \citet{Despali+2016} we find are: $A = 0.331, a=0.831, p=0.351$.  Fig. \ref{fig:despaliFit} shows the resulting HMF for several redshifts from 0 to 5.

Besides the variations due to different cosmological parameters \citep{Angulo+2010} it is difficult to model the uncertainties affecting the HMF.  
\citet{Despali+2016} thoroughly discuss the implications of different density thresholds in the spherical overdensity algorithm, e.g., replacing the virial overdensity  with 200 times critical ($\rho_\mathrm{c}$) or mean background ($\rho_\mathrm{b}$) density. They conclude that the virial definition leads to a  ``universal'' HMF fit 
while in the case of $\rho_\mathrm{c}$ or $\rho_\mathrm{b}$  the results are more redshift dependent. A higher density threshold -- e.g.~$500\rho_\mathrm{c}$, as often used in the literature -- alters the  HMF profile by decreasing the number density of the most massive systems, as some of them are now identified as a complex of smaller, independent haloes.
The assumption of sphericity in the finder algorithm is less problematic since its impact on the HMF is mass-independent: accounting for haloes' tri-axiality has only a mild effect on the HMF \citep{Despali+2014}.

Other studies have investigated the impact of different halo finding algorithms which produce changes in the HMF of the order of  $\sim10\%$ \citep{Knebe+2011}. Another potential issue is the impact of baryons (not implemented in \citeauthor{Despali+2016}) on the growth of dark matter haloes:  \citet{Bocquet+2016} show that in hydrodynamical simulations the halo number density decreases by $\sim$15\% at $z\lesssim0.5$ with respect to dark matter only, whereas at higher redshift the impact of baryons is negligible.

In our analysis we use \citeauthor{Despali+2016} mass function fitted on the Bolsho\"{i}-Planck simulation where haloes are identified using the virial overdensity criterion and where there mass is the maximal mass in their history \mhmax. We also use the original version of \citet{Despali+2016} HMF with halo mass defined with the virial overdensity criterion. To quantify how such a choice affects our results we consider different HMF estimates.
These alternate versions are divided into two categories according to how haloes are identified. HMF estimates in the first category \citep{Tinker+2008, Bocquet+2016} use the spherical overdensity definition, with halo masses defined with the $>200 \rho_\mathrm{b}$ criterion, while the others \citep{Bhattacharya+2011, Watson+2013} rely on the so-called ``friends of friends'' algorithm \citep{Davis+1985}. 
The  Bolsho\"{i}-Planck fit of \citeauthor{Despali+2016} HMF is shown in the upper panel of  Fig.\ \ref{fig:SMF&HMF} while the lower panel shows how this fit and the other HMF differ from \citeauthor{Despali+2016} in three redshift bins. At low redshifts we find that differences are negligible, in agreement with the literature. However for $>$10$^{13}\,M_\odot$ haloes at $z>2$, i.e.\ in a range barely investigated in previous work, there are $0.2\!-\!0.5$\,dex  offsets between \cite{Despali+2016} and other HMF estimates. Such a difference may be fully explained by Poisson scatter since such massive haloes are rare in the volume of cosmological simulations. We do not attempt to find the physical reasons of such a  discrepancy and here we simply take the ``inter-publication'' bias as a measure of generic HMF uncertainties (see Sect.\ \ref{subsec:HMFimpact}).

\section{The stellar-to-halo mass relationship}
\label{sec:SHMR}

\subsection{The sub-halo abundance matching technique}
\label{sec:TheorySHMR}

In the sub-halo abundance matching (SHAM) technique a ``marker'' quantity is assigned to dark matter haloes and galaxies (e.g.\ halo mass and stellar mass, respectively). Both haloes and galaxies are ranked according to their marker quantity, and then the latter are associated to the former by assuming a monotonic one-to-one relationship \citep{Vale+2004,Conroy+2006,Behroozi+2010,Moster+2010,Reddick+2013}. Here, the markers we use are the dark matter halo mass and the galaxy stellar mass.
 
In the hierarchical clustering scenario small haloes accrete onto more massive ones and become ``sub-haloes''. Galaxies are classified as either ``satellite'' (those hosted by sub-haloes) or ``central'' (those in the main halo). Since in the COSMOS2015 catalogue there is no distinction between satellite and central galaxies to correctly perform the abundance matching we must consider all the haloes (i.e., main haloes and sub-haloes) as a whole sample. For sake of simplicity, we will refer to any (sub-)halo hosting a galaxy as a ``halo''. We do not take into account possible ``orphan'' galaxies \citep[i.e., satellites with no sub-halo,  e.g.][]{Moster+2013}.

These orphan galaxies may appear when matching a catalogue of dark matter haloes with galaxies from observations \citep[this is done in e.g.][]{Moster+2018, Behroozi+2018}. If the resolution of the simulation (or of the halo finder) is not precise enough, the catalogue may miss the smallest haloes and some galaxies will be unassociated. In our work we do not use directly halo catalogues from a simulation but instead fits of a functional form of the HMF performed on outputs of simulations. \citet{Despali+2016} made sure that the fit is performed on a range of halo masses not affected by the limits of the simulation and the HMF is extrapolated below this limit to smaller masses. \citet{Campbell+2018} investigated the importance of the orphan galaxies in the Bolsho\"{i}-Planck simulation. They concluded that less than 1\% of galaxies with $M_* > 10^{9.5} h^{-2} M_\odot$ are orphans. As such we consider that the different HMF we use are not impacted by the resolution limits of the simulations and that the impact of orphan galaxies on the SHMR is negligible in the range of mass we consider.  

The SHAM method also does not consider either the gas mass or the intracluster medium. Our sources of uncertainties are discussed in more detail in Section \ref{sec:limit} \citep[see also][]{Behroozi+2010,Campbell+2018}. 

We perform a ``parametric'' SHAM, assuming a functional form for the  relation between $M_*$ and  $M_{\rm h}$. Following the same formalism as in \cite{Behroozi+2010}, such a parametric SHMR is described by the following equation:

\begin{equation}
\begin{split}
\log(M_{\rm h}) = \log(M_{1}) + \beta \log\left(M_{*}/M_{*,0}\right) + \\
+\frac{\left(M_{*}/M_{*,0}\right)^\delta}{1+\left(M_{*}/M_{*,0}\right)^{-\gamma}} - \frac{1}{2} \;.
\end{split}
\label{eq:SHMR}
\end{equation}
This model has five free parameters $M_{1}$, $M_{*,0}$, $\beta$, $\delta$, $\gamma$, which determine the amplitude, the shape and the knee of the SHMR \citep[see][for  a more detailed description of the role of each parameter in shaping the SHMR]{Behroozi+2010}.
Roughly speaking, parameter values in Eq.\ (\ref{eq:SHMR}) are adjusted during an iterative process until the HMF, converted into stellar mass through the SHMR, is in agreement with the observed SMF (see Sect. \ref{sec:MCMCfit}).

More specifically, the galaxy cumulative number density ($N_\ast$) and the halo cumulative number density ($N_{\rm h}$) above a certain mass are respectively given by $N_\ast (M_\ast) = \int_{M_\ast}^{+\infty} \phi_\ast(M)\mathrm{d}M$ and $N_{\rm h} (M_{\rm h}) = \int_{M_{\rm h}}^{+\infty} \phi_{\rm h}(M)\mathrm{d}M$, with $\phi_\ast$ and $\phi_{\rm h}$ being the stellar and halo mass functions. The main assumptions of SHAM is that there is only one galaxy per dark matter halo and that the relation between stellar and halo masses is monotonic. As a consequence, the $M_\ast$  value associated to a given $M_\mathrm{h}$ is the one for which $N_\ast (M_\ast) = N_{\rm h} (M_{\rm h})$.
The derivative of this equation gives the relationship between SMF, HMF, and SHMR:

\begin{equation}
\phi_\mathrm{\ast,conv}(M_\ast) = \frac{\mathrm{d} M_{\rm h}}{\mathrm{d}M_\ast} \phi_{\rm h}(M_{\rm h}) \,,
\label{eq:HMFtoSMF}
\end{equation}
where the differential term on the right-hand side can be derived from Eq.\ (\ref{eq:SHMR}). We use the notation $\phi_\mathrm{\ast, conv}$ because we convolve this SMF with a log-normal distribution to account for scatter in stellar mass at fixed halo mass.
The standard deviation ($\xi$) of the log-normal distribution is kept as an additional free parameter; we assume that  $\xi$ is independent of the halo mass  \citep{More+2009,Moster+2010} but can vary with redshift. We note here that new hydrodynamical simulations like Eagle \citep{Schaye+2015} have shown that this scatter decreases from 0.25 dex at $M_{\rm h} = 10^{11} M_\odot$ to 0.12 dex at $M_{\rm h} = 10^{13} M_\odot$ \citep[see][]{Matthee+2017}. This evolution of the scatter is in agreement with latest abundance matching models \citep{Coupon+2015, Behroozi+2018, Moster+2018}. See also figure 9 of \citet{Gozaliasl+2018}.
However in our analysis we restrict ourselves to a mass-invariant scatter for simplicity.

The model SMF defined in Eq.~(\ref{eq:HMFtoSMF}) is then fitted to the observed one (i.e., $\phi_\mathrm{\ast,obs}$) through the procedure described in the next Section.

\subsection{Fitting procedure}
\label{sec:MCMCfit}

To fit the model SMF to real data, a negative log-likelihood is defined as:

\begin{equation}
\chi^{2} = \sum_{i} \left(\frac{\phi_{\ast,\rm conv}(M_{*,i}) - \phi_{\rm \ast,obs}(M_{\ast, i})}{\sigma_{\rm obs}(M_{*,i})}\right)^{2},
\label{eq:chi2}
\end{equation}
where $\sigma_{\rm obs}$ is the uncertainty of the observed SMF  in a given stellar mass bin $M_{*,i}$ (with the first bin starting at $M_\mathrm{\ast,min}$).

For each of the ten redshift bins, we minimise Equation (\ref{eq:chi2}) using a Markov Chain Monte Carlo (MCMC) algorithm\footnote{We use the \textsc{Emcee} \texttt{python} package \citep{Foreman+2013}.}. This algorithm allow the sampling of the parameter space in order to derive the posterior distribution for the six free parameters. We use flat conservative priors on the parameters together with 250 walkers each with a different starting point randomly selected in a Gaussian distribution around the original starting point. Our convergence criterion is based on the autocorrelation length, which is an estimate of the number of steps between which two positions of the walkers are considered uncorrelated \citep{Goodman+2010}. Our MCMC stops when the autocorrelation length has changed by less than 1 per cent and when the length of the chain is at least 50 times the autocorrelation length. As an example, our chains in the case of the HMF fitted on Bolsho\"{i}-Planck have a length between 5000 at low redshift and 25000 in the highest redshift bin. With our 250 walkers this gives between $1.25\times10^{6}$ and $6.25\times10^{6}$ samples. The first steps up to two times the autocorrelation length are discarded as a burn-in phase. To speed up the computation of the posteriors we keep only the iterations separated by a thin length which is half of the autocorrelation length.

We show in Table \ref{tab:MCMC_fits} the best fit and the 68 per cent confidence interval for the six free parameters in each of the ten redshift bins, along with the marginalised posterior distributions in Fig. \ref{fig:post_0-1}, \ref{fig:post_2-5} and \ref{fig:post_6-9}.
These figures show that the parameters $M_1$ and $M_{*,0}$ are highly correlated. This is expected because as $M_1$ increases, $M_{*,0} $ should also increase. $M_1$ and $\beta$ are also highly correlated which may be explained by the fact that $\mathrm{log}(M_*/M_{*,0})$ is negative for a large range of stellar masses so an increase of $\beta$ is compensated by an increase of $M_1$. As we can see, the value of $\delta$ is not well constrained at high redshift, because this parameters controls the high mass slope which is not well constrained in our data.

\subsection{Principal sources of SHAM uncertainties}
\label{sec:limit}

There are several sources of uncertainties in the SHAM technique. A sub-halo may be stripped after infall, leaving the hosted galaxy embedded in the larger, central halo. This may break the one-to-one correspondence between galaxies and dark matter haloes which is the main assumption of our method. The HOD model is a viable solution to take this into account although it would introduce an additional number of assumptions and free parameters. Moreover, observed galaxy clustering is required to constrain the HOD model parameters \cite[e.g.][]{Coupon+2015} but such measurements are  challenging at $z>2$ \citep{Durkalec2015}. At lower redshift ($z\lesssim1$) \citet[][see their fig. 13]{Leauthaud+2012} have shown that \mhp measurements are consistent between HOD and SHAM measurements.  

Another source of uncertainty comes from random and systematic errors in $z_\mathrm{phot}$ and $M_\ast$ estimates, with the former propagating into the $M_\ast$ error in a way difficult to model (see discussion in D17).  In D17 the logarithmic stellar mass uncertainty is described by a Gaussian with standard deviation $\sigma_{M\ast}=0.35$\,dex multiplied by a Lorentzian function with a parameter $\tau$ increasing with redshift to enhance the tails of the distribution (see equation 1 of D17). These observational uncertainties which cause the Eddington bias have been corrected for in the SMF estimates we adopt (Section \ref{sec:MassFunctions}) but some caveats remain \citep[see D17;][]{Grazian+2015}.
Moreover, in our fitting procedure, we consider that different stellar mass bins are uncorrelated (Eq.\ \ref{eq:chi2}). This assumption is a consequence of the fact that in D17 (as the vast majority of the literature) covariance matrices are not provided for their SMF estimates. Once corrected for $M_\ast$ \textit{observational} uncertainties, the main source of correlations between mass bins comes from the \textit{intrinsic} covariance between them. 
To avoid oversampling, we adopt a mass bin size of 0.3\,dex which is comparable to the  scatter in D17. 
We verified that this choice does not introduce any significant bias: modifying the bin size and centre (by $\pm0.1$\,dex) results remain consistent within 1$\sigma$. 

Besides their impact on $M_\ast$ estimates, $z_\mathrm{phot}$ uncertainties affect the observed SMF by scattering galaxies in the wrong redshift bin. Our binning is large enough to mitagate this given that typical $z_\mathrm{phot}$  dispersion in COSMOS2015 estimated from a large spectroscopic galaxy sample  reaches  $\sigma_z \simeq 0.03(1+z)$ at $2.5<z<6$.

Nonetheless, catastrophic $z_\mathrm{phot}$ errors in the SED fitting (e.g., due to degenerate low- and high-$z$ solution) may still be a concern. The fraction of catastrophic redshift outliers in COSMOS2015 is about $0.5\%$ at $z<3$ and $12\%$ at higher redshifts, so it should not introduce a significant covariance between $z$ bins. This seems to be confirmed by test with hydrodynamical simulations (Laigle et al., in prep.).

Despite this, the impact of SED fitting systematics is still an open question which will only be resolved with next-generation surveys (e.g., large and unbiased spectroscopic samples with the Prime Focus Spectrograph at Subaru, or the \emph{James Webb} Space Telescope).

However, in this work we independently constrain parameters of Eq.\ (\ref{eq:SHMR}) at each redshift bin without assuming a functional form for their redshift  evolution \citep[contrary to][]{Behroozi+2010}. Such a redshift--independent fit reduces the overall number of assumptions in the SHMR modelling.

A final source of uncertainty comes from the $M_\ast$ scatter we add to the galaxy-to-halo monotonic relation. This is modelled with a log-normal distribution characterised by the parameter $\xi$  which is free to vary in the MCMC fit. This parameter is usually fixed between 0.15 and 0.20 dex \citep[see e.g.][]{More+2009, Moster+2010, Reddick+2013} but in the large redshift range probed here we expect a non-negligible variation due to the evolution of galaxies' physical properties as well as observational effects. We note however that the resulting values (see Table \ref{tab:MCMC_fits}) are compatible with the fixed ones assumed by the studies mentioned previously.

\section{Results}
\label{sec:Results}

\subsection{The stellar-to-halo mass relationship}
\label{sec:ResultsSHMR}

\begin{figure}
  \includegraphics[width=\columnwidth]{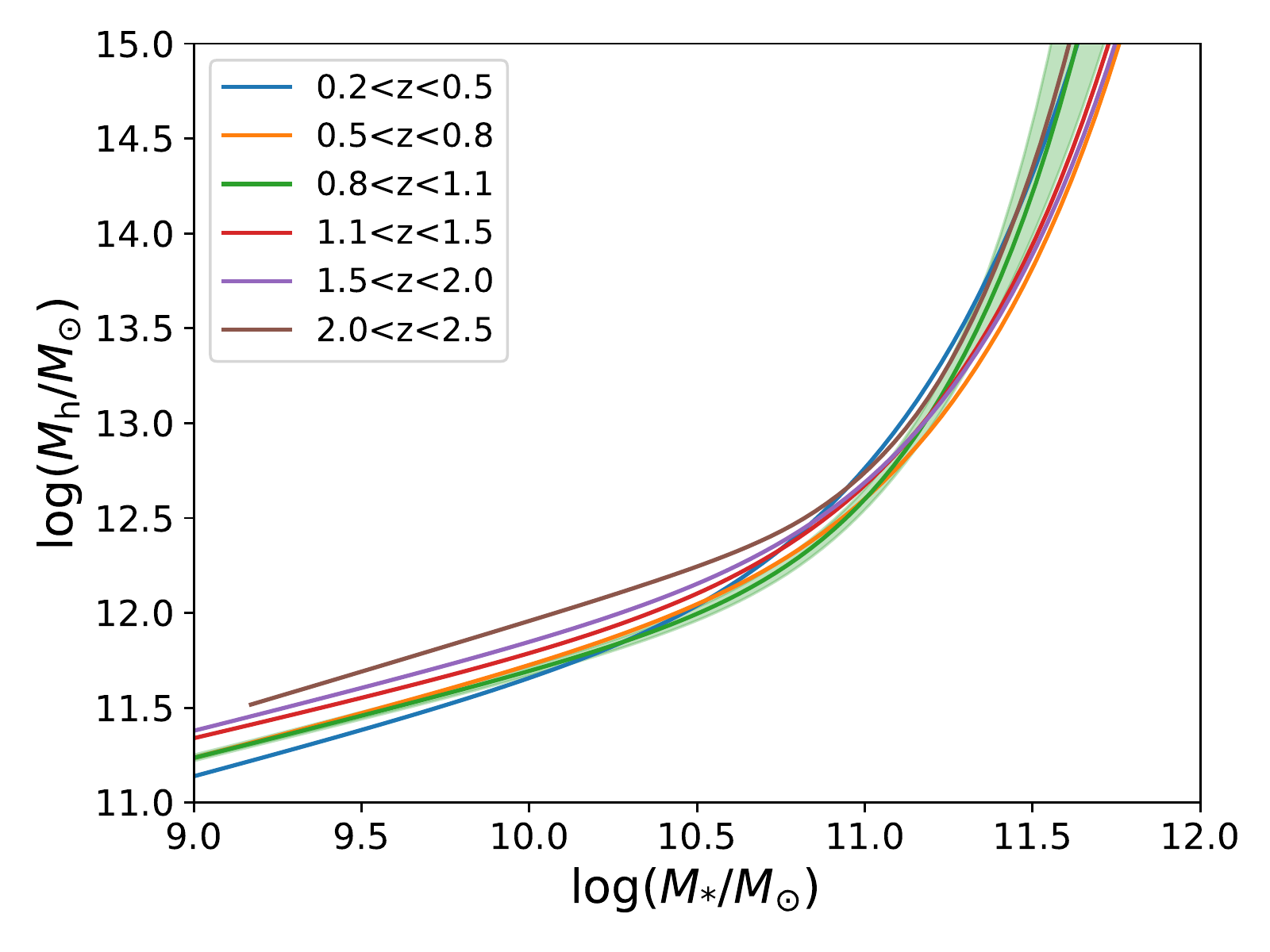}\\
   \vspace{5pt}
\includegraphics[width=\columnwidth]{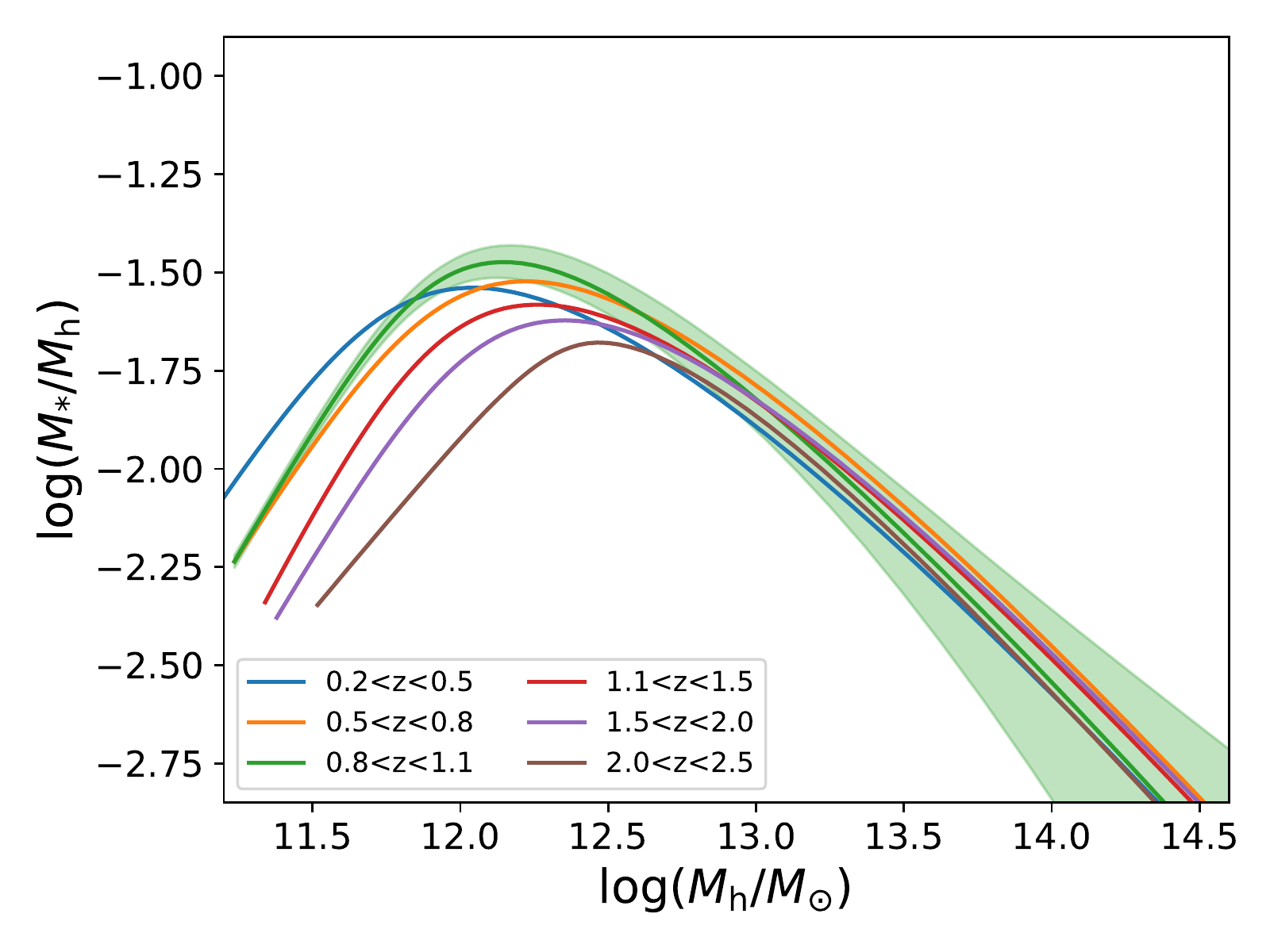}
  \caption{ Upper panel: Stellar-to-halo mass relation from $z=0.2$ to $z=2.5$. Thick lines show the 50\textsuperscript{th} percentile of the $M_{\rm h}$ distribution at fixed $M_*$ computed from our MCMC runs. The coloured bands show the 16\textsuperscript{th} and 84\textsuperscript{th} percentile. The band is shown for the $0.8<z<1.1$ redshift bin only for clarity (for other redshift bins, the uncertainty is of the same order). The limits in stellar mass for each redshift is derived from observations. Lower panel: \msmh ratio derived from this SHMR.}
  \label{fig:SHMRCosmos_upto6}
\end{figure}

\begin{figure}
  \centering
 \includegraphics[width=\columnwidth]{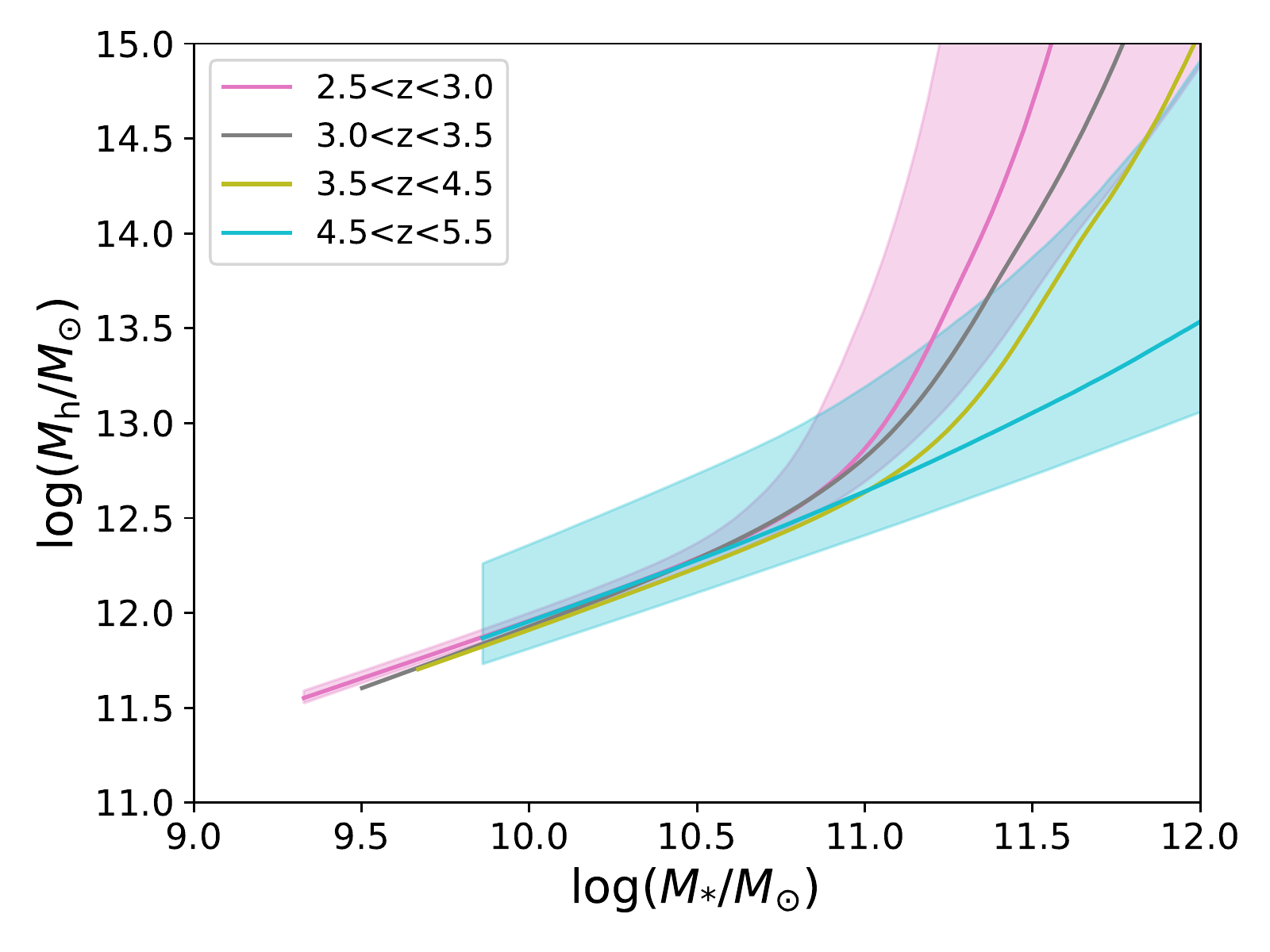}\\
  \vspace{5pt}
\includegraphics[width=\columnwidth]{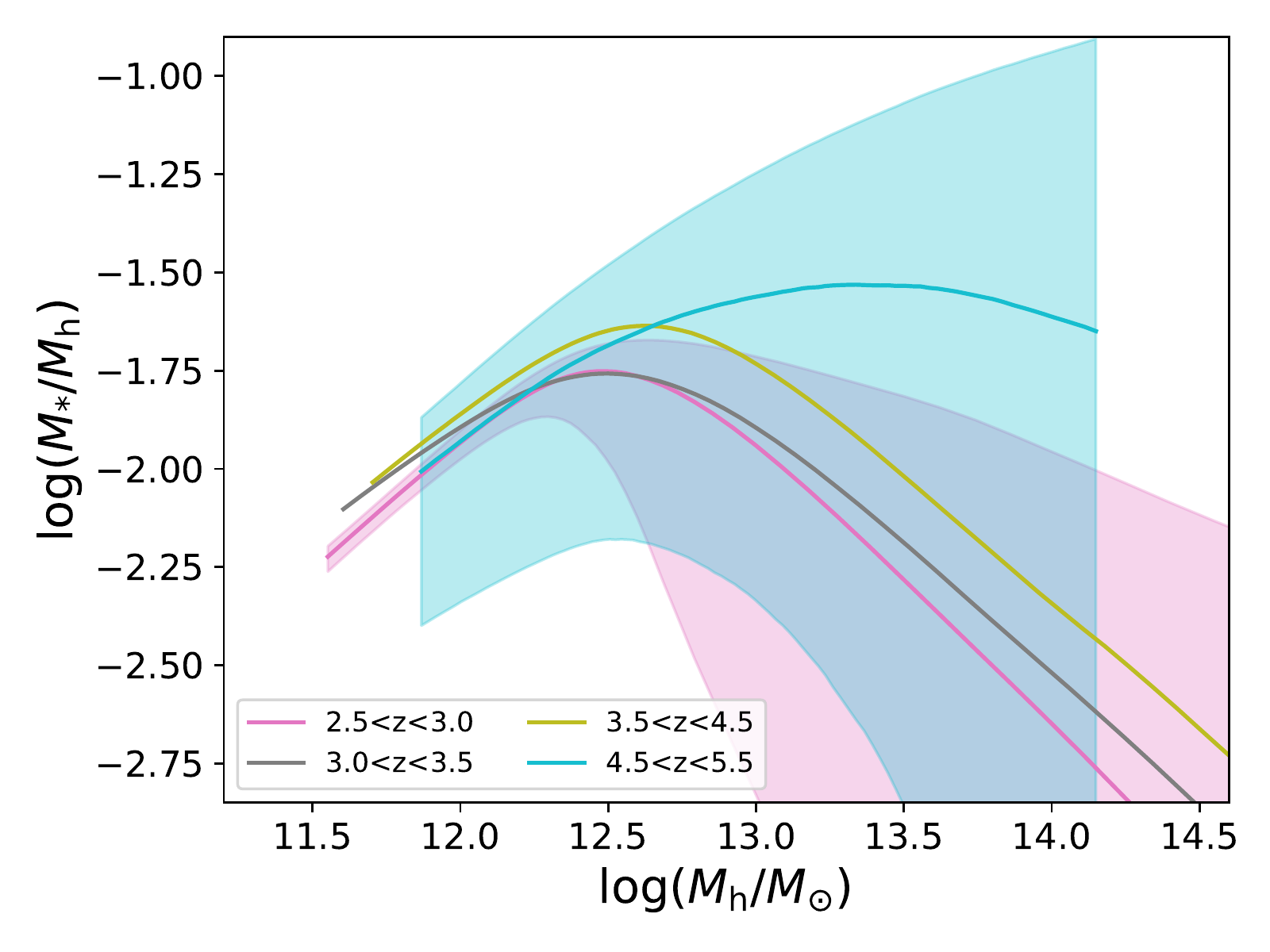}
  \caption{Same as Fig. \ref{fig:SHMRCosmos_upto6} for redshift bins from $z=2.5$ to $z=5.5$. We show only uncertainties for the $2.5<z<3.0$ and $4.5<z<5.5$ bins for clarity.}
  \label{fig:SHMRCosmos_789}
\end{figure}

Fig.~\ref{fig:SHMRCosmos_upto6} and \ref{fig:SHMRCosmos_789} show our derived SHMR fits (upper panels) and the corresponding ratio between stellar mass and halo mass (lower panels) for samples in $0.2<z<2.5$ and $2.5<z<5.5$ redshift intervals respectively. The SHMR and the corresponding $1\sigma$ uncertainty are computed respectively as the 50\textsuperscript{th}, the 16\textsuperscript{th} and the 84\textsuperscript{th} percentile of the distribution of $M_{\rm h}$ at a given $M_{*}$ in the remaining MCMC chains (Section~\ref{sec:MCMCfit}).  These uncertainties are shown as the shaded regions. Considering the stellar  mass completeness of our dataset (Section \ref{sec:MassFunctions}) we limit our samples to $M_\ast>M_{*,\rm min}(z)$ and also restrict ourselves at $M_\mathrm{h}<10^{15} M_\odot$ since the number density of haloes of such a mass is negligible across the whole redshift range \citep[$<10^{-6}\mathrm{Mpc^{-3}}$; see][]{Mo&White2002}).

We note that the stellar mass evolution of haloes between redshift bins might sometime appear at odds with the expected stellar mass assembly. A halo with $M_\mathrm{h} = 10^{13} M_\odot$ has a stellar mass of $10^{11.27} M_\odot$ at $z=4$. This halo is expected to grow to a mass of $M_\mathrm{h} = 10^{13.5} M_\odot$ at $z=2.5$ where our model says the galaxy should have a stellar mass of $10^{11.12} M_\odot$. This effect of haloes ``loosing'' stellar mass is a consequence of the fact that in our analysis each redshift bin is treated independently and the consistency of the model across different epochs is not guaranteed. The offset of about -0.15 dex in the example above probably arises from systematic uncertainties in the stellar mass function at high redshift, from SED fitting effects \citep[see e.g. discussion in ][]{Stefanon+2015} or from cosmic variance issues\citep[see e.g.][]{Davidzon+2017}. Besides systematic errors, statistical uncertainties are already able to explain part of the issue: running our MCMC with a SMF shifted by $-1 \sigma$ (statistical error) at $z=4$, and $+1 \sigma$ at $z=2.5$, the stellar mass difference in the example above is only $-0.06 \mathrm{dex}$.

The SHMR in the various redshift bins (upper panels of Fig.~\ref{fig:SHMRCosmos_upto6} and \ref{fig:SHMRCosmos_789}) monotonically increases as a function of stellar mass with a changing of slope at $\sim$\mhpp. Below the characteristic halo mass, the SHMR slope is approximately constant with redshift. Conversely, for masses above \mhpp, it becomes flatter as moving towards higher redshifts (Figure~\ref{fig:SHMRCosmos_789}, upper panel) modulo the large error bars especially at $4.5<z<5.5$.
 
These trends are clearly illustrated also in the lower panels of Fig.~\ref{fig:SHMRCosmos_upto6} and \ref{fig:SHMRCosmos_789} which show \msmhp vs $M_h$. In each bin, this ratio peaks at  $M_\mathrm{h}\simeq$ \mhp and drops by one order of magnitude at both the extremes of our halo mass range. At $z<0.5$, \mhp $= 10^{12} M_{\odot}$, with $\mathrm{log}(M_{*}/M_{\rm h}) = -1.55 \pm 0.5$. At higher redshifts, \mhp increases steadily up to $10^{12.5} M_{\odot}$ at $z=2$, i.e.\ growing by a factor $\sim$3. It then remains flat up to $z=4$. At a fixed halo mass above \mhpp, \msmh does not evolve, while in haloes below \mhp the ratio decreases from $z\sim0$ to $z\sim2.5$.

\subsection{Dependence of the peak halo mass on redshift}
\label{sec:peak}

\begin{figure*}
\includegraphics[width=2\columnwidth]{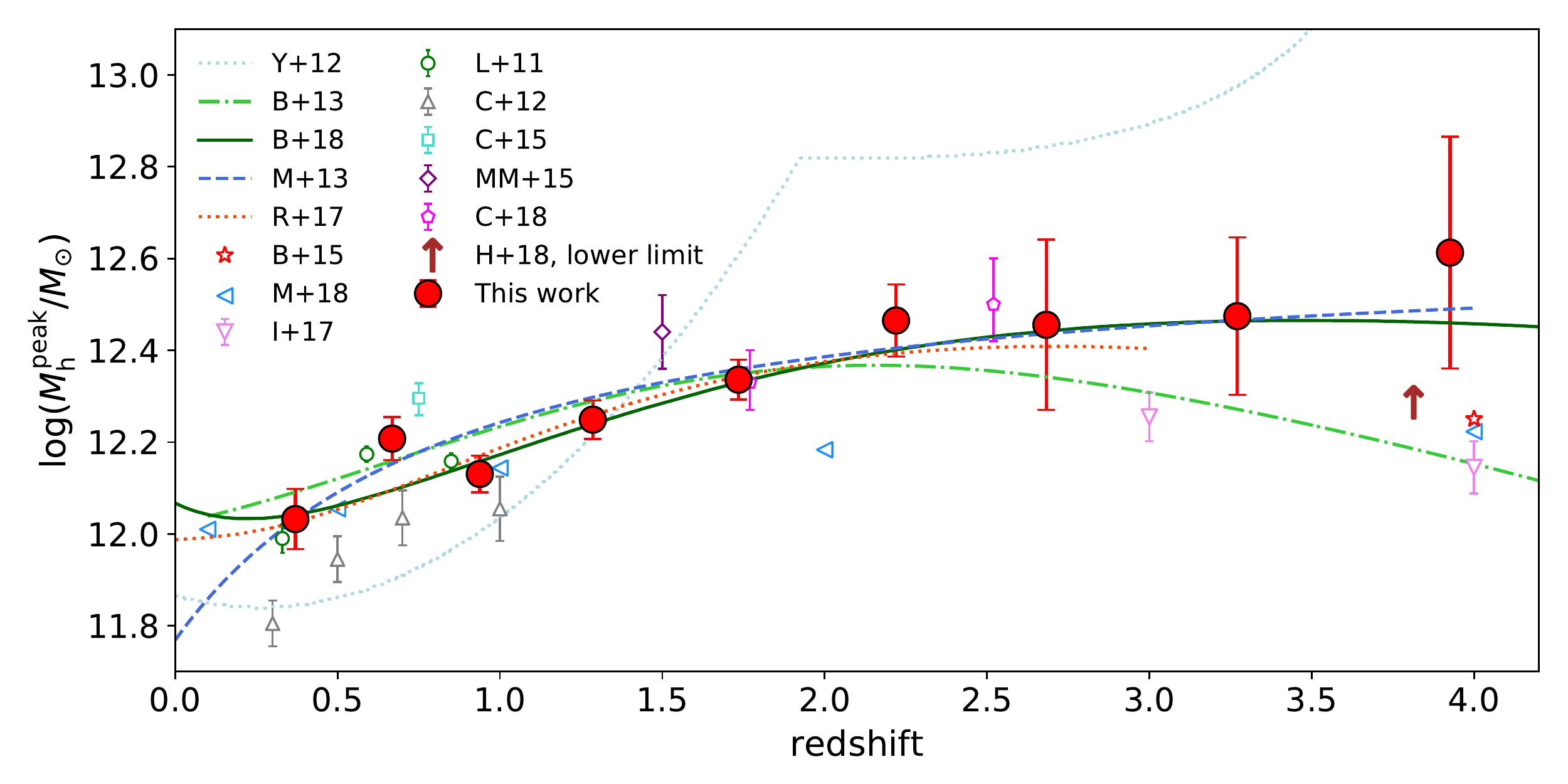}
\caption{ Peak halo mass \mhp as a function of redshift (red dots). We plot \mhp at the median redshift of each bin, rescaled to $H_{0} = 70\ \rm km.s^{-1}Mpc^{-1}$. All masses from other studies have been rescaled to match the $H_{0} = 70 \ \rm km.s^{-1}Mpc^{-1}$ cosmology. Some points from the literature have been slightly shifted along the redshift axis for clarity. We show results from \citet[][L+11]{Leauthaud+2011}, \citet[][Y+12]{Yang+2012}, \citet[][C+12]{Coupon+2012}, \citet[][M+13]{Moster+2013}, \citet[][B+13]{Behroozi+2013_SHMR}, \citet[][B+15]{BehrooziSilk2015}, \citet[][C+15]{Coupon+2015}, \citet[][MM+15]{Martinez-Manso+2015}, \citet[][R+17]{Rodriguez-Puebla+2017}, \citet[][I+17]{Ishikawa+2017}, \citet[][C+18]{Cowley+2018}, \citet[][H+18]{Harikane+2018}, \citet[][M+18]{Moster+2018}, and \citet[][B+18]{Behroozi+2018}. The brown arrow is the lower limit for \mhp from \citet[][H+18]{Harikane+2018}.}
\label{fig:MhaloPeak_z}
\end{figure*}

Fig.~\ref{fig:MhaloPeak_z} shows the redshift evolution of the peak halo mass between $z=0.2$ and $4.5$ computed from the median \mhp for all the samples retained in the MCMC (see Section \ref{sec:MCMCfit}). The results are reported also in Table \ref{tab:MCMC_fits}. Fig.~\ref{fig:MhaloPeak_z} also presents a compilation of recent measurements from the literature together with model predictions (lines). At $z>3$ it becomes progressively more difficult to measure the position of the peak as the slopes of halo  and stellar mass functions become similar (Fig.\ref{fig:SMF&HMF}). In addition at higher redshifts there are correspondingly smaller numbers of massive galaxies in the COSMOS volume. Nevertheless, our measurements show clearly that the peak halo mass increases steadily from $10^{12} M_\odot$ at $z=0.3$ to $10^{12.6} M_\odot$ at $z=4$. 

Below $z\sim2.5$ there is generally a good agreement in the literature with \mhp  steadily increasing as a function of redshift. We confirm this trend despite some fluctuation (e.g., at $z\sim0.7$) due to the over-abundance of rich structures in COSMOS \citep[see e.g.][]{McCracken+15}. \citet{Leauthaud+2012}, using a previous measurement of the COSMOS SMF at $z<1$, find the same fluctuations. \citeauthor{Leauthaud+2012} perform a joint analysis of galaxy-galaxy weak lensing and galaxy clustering to fit the SHMR modelled as in  \citet{Behroozi+2010}. Moreover, they use a halo occupation distribution to describe the number of galaxies per dark matter halo, instead of assuming only a single galaxy inhabits each dark matter halo. In fact, such an assumption has only a small impact on the \mhp position given the fact that at $\sim$10$^{12}\,M_{\odot}$ most of the haloes contain only one galaxy \citep{McCracken+15}. 
\citet{Cowley+2018} used an HOD model to derive \mhp for mass-selected sample of UltraVISTA galaxies in COSMOS at $1.5<z<2$ and $2<z<3$; their results are in good agreement with ours. Their error bars account for $z_\mathrm{phot}$ errors but not the stellar mass uncertainties; in their HMF, they apply \citet{Behroozi+2013_Rockstar} high-redshift correction and introduce a large-scale halo bias parameter \citep{Tinker+2010}. 

Above $z\gtrsim3$ the scatter in \mhp  increases. \citet{Moster+2013} and \citet{Behroozi+2013_SHMR}\footnote{Values shown here were obtained using Planck cosmology instead of the published WMAP cosmology (P. Behroozi private communication). See comparison in fig. 35 of \citet{Behroozi+2018}.} find different trends, i.e.\ a   \mhp$(z)$ function that declines \citep{Behroozi+2013_SHMR} or flattens \citep{Moster+2013} with increasing redshift. One possible explanation for the discrepancy is that \citeauthor{Moster+2013} and \citeauthor{Behroozi+2013_SHMR} models are based on different observational datasets. To address this issue, \linebreak \citet{BehrooziSilk2015} repeated \citeauthor{Behroozi+2013_SHMR}'s analysis removing $z>5$ constraints (which in their method influence also the fit at lower $z$). However, this test is inconclusive as their \mhp estimate (shown as the star symbol in Fig.~\ref{fig:MhaloPeak_z}) falls between these curves\footnote{
 \mhp$(z)$ error bars are not explicitly quoted either in \citet{Behroozi+2013_SHMR} or \citet{Moster+2013}. However, we can quantify them through the  uncertainties of their SHMR models. For example in the model of \citeauthor{Moster+2013} the 1$\sigma$ confidence level of the $M_1(z)$ parameter can be used as a proxy, leading to \mhp error bars of the same order of magnitude of ours.}.

Our higher \mhp values with respect to \citet{Ishikawa+2017} and \citet{Harikane+2018} may be a consequence of our near-infrared selection (a good proxy for stellar mass, see D17). \citet{Ishikawa+2017} and \citet{Harikane+2018} samples are selected in rest-frame UV (and a conversion to stellar mass is made through an average $L_\mathrm{UV}$-$M_\ast$ relation). Moreover their redshift classification is derived  (instead  of $z_\mathrm{phot}$ estimates) from a Lyman-break colour--colour selection which may result in lower levels of purity and completeness at $z\sim3$ \citep{Duncan+2014}.
  
Recently, revised versions of \citet[][]{Behroozi+2013_SHMR} and of \citet[][]{Moster+2013} have been presented in \citet[][]{Behroozi+2018} and \citet{Moster+2018}. This new analysis differs from the former ones by following closely the evolution of individual halo-galaxy pairs through time. This results in a better understanding of the scatter of the SHMR, because this scatter results from the evolution of each halo-galaxy pairs, it is not an arbitrary scatter parameter added to the model.
In \citet{Behroozi+2018}, the feedback model regulating star formation has significantly changed since \citet{Behroozi+2013_SHMR}. In the updated model, the $M_\mathrm{h}$ threshold at which 50 per cent of the hosting galaxies are quiescent grows from $10^{12}\,M_\odot$ at $z<1$ up to $\sim$10$^{13}\,M_\odot$ at $z=3.5$ \citep[see Fig.~28 of][]{Behroozi+2018}. As a consequence, the \mhp evolution is now in excellent agreement with both \citet{Moster+2013} and our estimates. 
\citet{Moster+2018} peak halo mass shown here corresponds to the peak in the ratio between stellar mass and baryonic mass of galaxies (the $\left[M_* / M_b \right] (M_h)$ relation. We assumed here that the ratio between baryonic mass and halo mass is a constant (equal to the universal baryon fraction), giving the same value for the peak halo mass of the $\left[M_* / M_h \right] (M_h)$ relation. The difference with our results might be explained by a dependence of the baryon fraction of haloes with mass \citep[see][]{Kravtsov+2005, Davies+2018}.

\subsection{Dependence of \msmh on redshift at fixed  halo mass}
\label{sec:MsMh_ratio}

\begin{figure}
\includegraphics[width=\columnwidth]{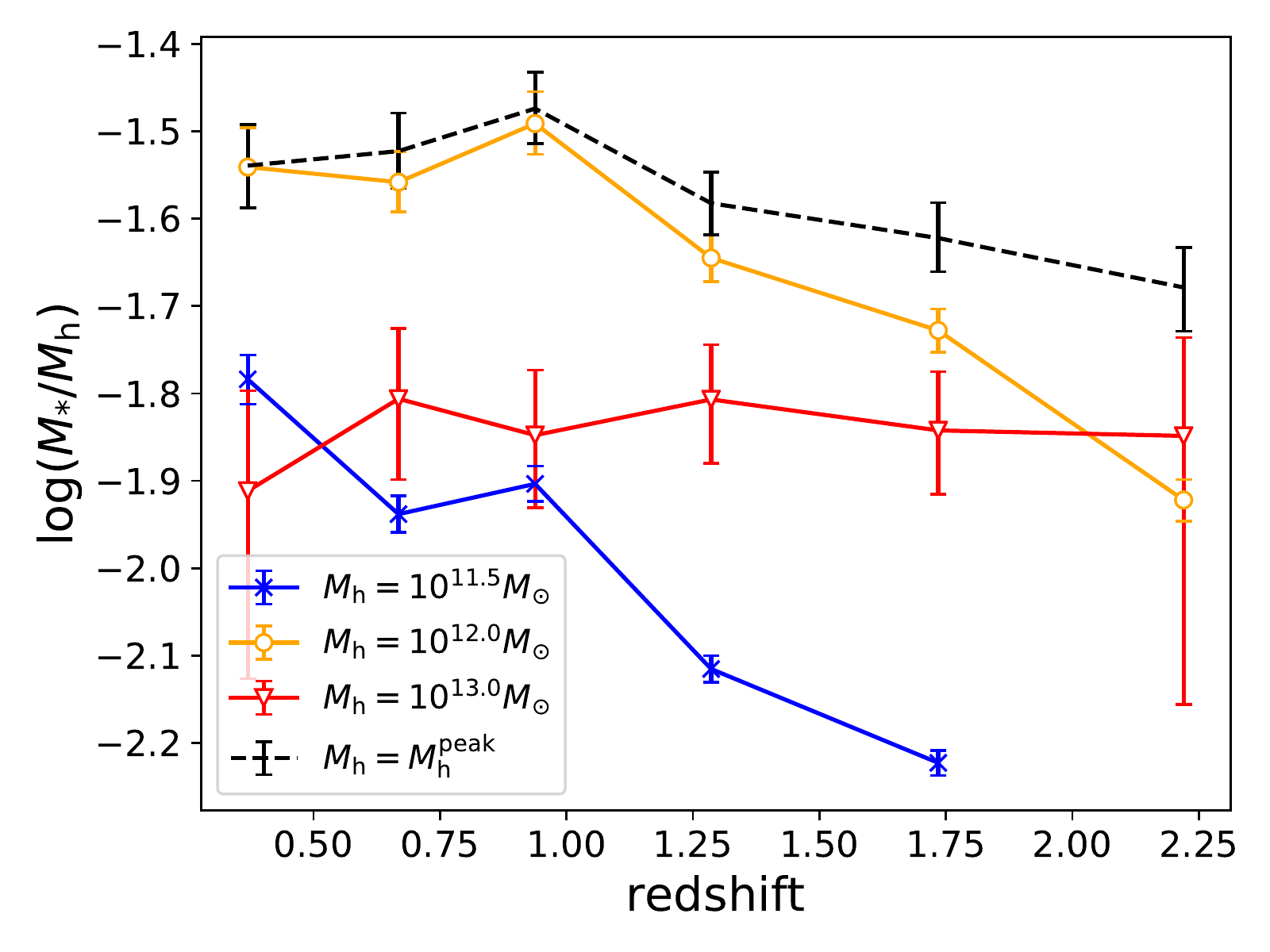}
\caption{Evolution of \msmh as a function of redshift for fixed halo masses (solid lines) and at $M_\mathrm{h}\,\equiv$\,\mhp (dashed line). Error bars are derived from Fig.~\ref{fig:SHMRCosmos_upto6} and \ref{fig:SHMRCosmos_789}.  
}
\label{fig:MsMh_fixMh}
\end{figure}

Since \msmh depends on host halo mass and redshift, we show in Fig.~\ref{fig:MsMh_fixMh} this trend in more detail by computing the \msmh ratio at different fixed values of halo mass. We restrict our analysis to $z<2.5$ because at high mass mass bins ($10^{13} M_\odot$) uncertainties in the \msmh ratio prohibits a quantitative discussion of its evolution with redshift between $z=2.5$ and $z=5.5$.

For massive haloes ($10^{13} M_{\odot}$) the ratio is nearly constant between $z\sim0.2$ and $2.5$ whereas at $M_\mathrm{h}\simeq10^{12} M_{\odot}$ it increases as cosmic time goes by reaching the maximum value (about $0.03$) at $z\simeq1$ and then remaining constant until $z\sim0.2$. The redshift at which $10^{12} M_\odot$ haloes reach the maximum \msmh ratio corresponds to the epoch when \mhp is equal to their mass.  Lower-mass haloes, which are $<$\mhp across the whole redshift range, steadily increase their \msmh  without any peak or plateau. For instance haloes with $M_\mathrm{h}\simeq10^{11.5} M_\odot$ increase their  \msmh ratio by a factor $\sim3.2$ from $z=2.5$ to 0.2. 
For comparison, Fig.~\ref{fig:MsMh_fixMh} also shows the increase of the \msmh ratio, from $z=2.5$ to 0.2, for haloes in a mass bin that evolves with redshift, i.e.\ $M_\mathrm{h}=M_\mathrm{h}^\mathrm{peak}(z)$.  
We discuss in Section \ref{sec:Discussion} the interpretation of these evolutionary trends and the implications in terms of galaxy star formation efficiency.  

\subsection{Impact of halo mass function uncertainties}
\label{subsec:HMFimpact}

\begin{figure}
\includegraphics[width=\columnwidth]{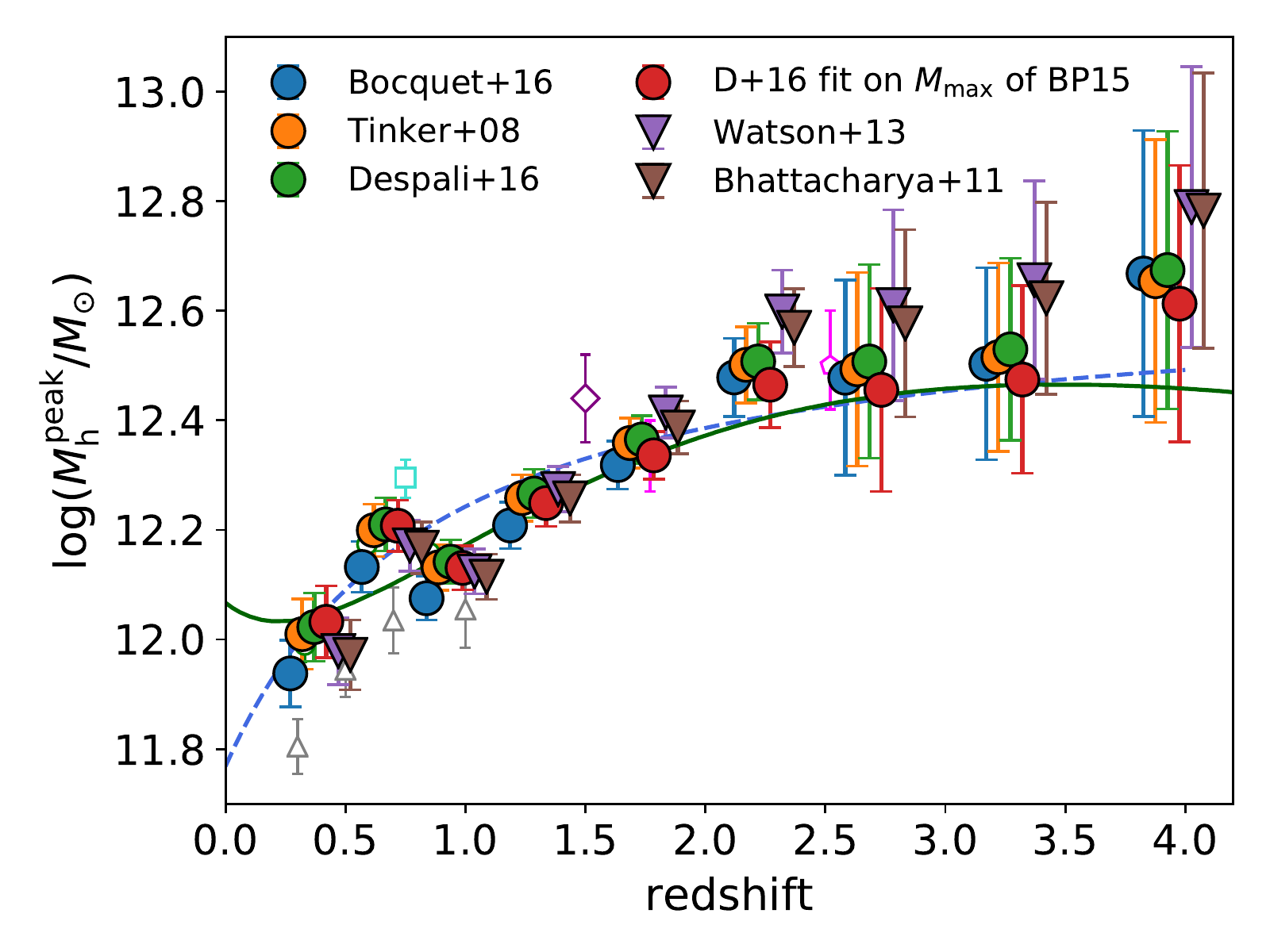}
\caption{ Peak halo mass (\mhpp) computed using different HMFs.
 \mhp redshift evolution is independently measured six times, using different HMF fits: our Bolsho\"{i}-Planck fit of \citet[][our main reference also shown in Fig.~\ref{fig:MhaloPeak_z}]{Despali+2016}; the original \citet[][]{Despali+2016}; \citet{Tinker+2008,Bhattacharya+2011,Watson+2013,Bocquet+2016}. Filled circles (triangles) indicate that the halo identification has been done with a spherical overdensity (friends-of-friends) algorithm. Each set of \mhpp$(z)$ values derived for a given HMF is shifted by 0.05 in redshift for sake of clarity. SHAM method and observed SMF are the same for all estimates. Literature measurements are shown as in Fig. \ref{fig:MhaloPeak_z}.
}
\label{fig:MhaloPeak_HMF}
\end{figure}

In order to estimate quantitatively how the choice of the HMF fit impacts our results, we repeat our analysis (Section \ref{sec:SHMR}) using different HMFs (Fig.~\ref{fig:MhaloPeak_HMF}). The SMF remains D17 in all the cases. Results at $z\lesssim2$ are consistent, whilst at higher redshifts we clearly observe the impact of halo identification techniques. \mhp values using the HMF of  either \citet{Tinker+2008}, \citet{Bocquet+2016}, or \citet{Despali+2016} are grouped together, as those studies all applied a spherical overdensity  criterion to define haloes. \citet{Bhattacharya+2011} and \citet{Watson+2013} use a friends-of-friends algorithm, and the resulting $\log(M_\mathrm{h}^\mathrm{peak}/M_\odot)$ is systematically higher by $\sim$\text{0.1 dex} at $z \geqslant 2$. In our study, these differences are smaller than other sources of uncertainty, but it is clear that in future larger surveys these differences may become important. 

\section{Discussion}
\label{sec:Discussion}

\subsection{Evolution of the SHMR observed in COSMOS}

To interpret our results it is worth first recalling how the shape of the SMF changes from $z=5$ to $z=0$ (Fig.~\ref{fig:SMF&HMF}).
The number density of intermediate-mass galaxies ($10^{9.5} M_{\odot}<M_\ast<10^{11} M_{\odot}$) increases more rapidly compared to lower and higher masses galaxies. This causes the ``knee'' of the SMF at $M_\ast \sim 10^{11} M_{\odot}$ to become progressively more pronounced. In comparison halo number density evolution is nearly independent of mass so the shape of the HMF is similar between $z=2$ and $z=6$ (modulo a normalisation factor, see Fig.~\ref{fig:SMF&HMF}). The relative evolution of these two functions causes the changes in the \msmh ratio.

The redshift evolution of the SMF shape is governed by several factors. On one hand, towards higher redshifts the high mass end becomes increasingly affected by larger observational uncertainties \citep[especially photometric redshift catastrophic failures:][]{Caputi+2015,Grazian+2015}.  
At the same time, specific physical processes control star formation around the knee of the SMF which are different from those affecting galaxies at lower masses \citep{Peng+2010}. Here, we assume that most of the observational errors have been accounted for (Sect.~\ref{sec:MassFunctions}) and consequently the redshift evolution of \mhp we measure in COSMOS is primarily driven by physical mechanisms. 

The \msmh ratio is usually interpreted as the comparison between the amount of star formation and  dark matter accretion integrated over a halo's lifetime. Thus, a high \msmh ratio in a given $M_\mathrm{h}$ bin implies that those  haloes have been (on average) particularly efficient in forming stars.  
``Star formation efficiency'' is used hereafter to indicate ``galaxy formation efficiency'', i.e.\ the whole process of stellar mass assembly from baryon accretion to the collapse of molecular clouds inside the galaxy. In addition to the \textit{in situ} star formation, further stellar mass can be accreted via galaxy merging.  In such a framework the dependence of the \msmh ratio on halo mass and redshift can be explained by a combination of physical phenomena.   
Our observational constraint on \mhp can help to understand which mechanisms, amongst those proposed in the literature, are most responsible for regulating galaxy stellar mass assembly.  

\mhp can be considered as the threshold above which haloes maintain a nearly constant \msmh ratio across time (Fig.~\ref{fig:MsMh_fixMh}). At a fixed halo mass below \mhp, the \msmh ratio increases as comic time goes by, indicating that stellar mass has ``kept up'' with dark matter accretion. For a fixed halo mass above $M_\mathrm{h}\simeq M_\mathrm{h}^\mathrm{peak}$, host galaxies are more likely to enter in a quiescent phase  (``quenching'' the star formation) and thereafter passively evolve. Fig.~\ref{fig:MsMh_fixMh} clearly shows this evolution with redshift for fixed halo masses. For objects with $M_\mathrm{h}=10^{12}\,M_\odot$, their \msmh increases until $z\sim1$ (i.e., when $M_\mathrm{h}^\mathrm{peak}=1.3\!-\!1.6\times10^{12}\,M_\odot$) after which the ratio remains constant until $z=0.2$. Note that we do not track the evolution of individual haloes but instead the evolution of the \msmh ratio for a given halo mass. This makes the interpretation of the evolution of individual haloes with time more difficult (haloes at high redshift are not necessarily the same as haloes of the same mass at low redshift).
%(this is similar to the difference between Lagrangian and Eulerian coordinates in fluid mechanics). 

\subsection{What physical mechanisms regulate star formation in our sample?}

In this work we consider primarily the redshift evolution of \mhpp. Our deep near-IR observations allow us to leverage the COSMOS2015 galaxy sample to constrain that threshold up to $z=4$ (Fig.~\ref{fig:MhaloPeak_z}). We find that the \mhpp$(z)$ function changes slope at $z\sim2$, showing a plateau at higher redshift. This implies that the threshold for massive galaxies to enter the quenching phase depends on redshift: in the early universe quenching mechanisms are less effective for galaxies in haloes between $10^{12}$ and $10^{12.5}\,M_\odot$.  
This scenario should also take into account the contribution of major and minor mergers but in the redshift and halo mass ranges of  our analysis they can be considered sub-dominant (see \citealp{Davidzon+2018} and references therein). Therefore, in the following we will focus on quenching models affecting the \textit{in situ} star formation. 

In their cosmological hydrodynamical simulations, \citet{GaborDave2015} implement a heuristic  prescription  to halt star formation in systems with a large fraction of hot gas.\footnote{
Namely, their code prevents gas cooling in FoF structures by setting the circumgalactic gas equal to the virial temperature. This condition is triggered when a structure has 60 per cent of gas particles with a temperature  $>10^{5.4}$\,K \citep{Keres2005}.} 
The condition to trigger the quenching phase, which \citeauthor{GaborDave2015} call ``hot halo'' mode,  happens exclusively at $M_{\rm h} >  10^{12}\,M_\odot$ in their simulation. This halo mass threshold is in agreement with \mhpp. 
However, \citet{GaborDave2015} carry out their analysis at $z<2.5$. At higher redshift this prescription would be in disagreement with our results. Also \citet{Behroozi+2018}, considering the evolution of the quiescent galaxy fraction,  emphasise  that  a quenching recipe with a constant  temperature threshold could not explain the observational trend. As the difference between a constant and a time-evolving threshold becomes more relevant in the first $\sim$2\,Gyr after the Big Bang \citep[see figure 28 in][]{Behroozi+2018} our results are extremely useful to discriminate between these different scenarios. 

The hot halo model is agnostic regarding the sub-grid physics of the simulation: gas heating can be caused by either stable virial shocks  
\citep[][]{BirnboimDekel2003} or AGN feedback \citep[see a review in][]{HeckmanBest2014}. 
With respect to the former mechanism, simulations in \citet{DekelBirnboim2006} show that shock heating in massive haloes becomes inefficient at high redshift  because cold streams are still able to penetrate into  the system and fuel star formation \citep[see also][]{Dekel+2009}. However, despite that this trend is in general  agreement with our results there are  quantitative  differences in the evolutionary trend. With the fiducial parameters assumed in \citet{DekelBirnboim2006} the ``critical redshift'' at which $\sim$10$^{12}\,M_\odot$ haloes start to form stars more efficiently is $z_\mathrm{crit}\simeq1.5$. Moreover, according to their model \mhp should keep increasing at $z>z_\mathrm{crit}$ instead of plateauing.    

Quenching models more compatible with our observational results have been presented e.g.~in \citet{FeldmannMayer2015}. Under the assumption that gas inflow (thus star formation) is strongly correlated to dark matter accretion, the authors note that at $z>2$ massive haloes are still  collapsing fast and dark matter filaments efficiently funnel cool gas into the galaxy. At $z\lesssim2$ those haloes should enter in a phase of slower accretion that eventually impedes star formation by gas starvation. However, we caution that they study single galaxies in cosmological  zoom-in simulations: a larger sample may show considerable variance in the redshift marking the transition between the two dark matter accretion phases. 
In addition, we emphasise that not only the accretion rate  but also the cooling timescale is a strong function of redshift. Gas density follows the overall matter density of the universe, evolving as $\propto(1+z)^3$. Since the post-shock cooling time is proportional to gas density, it  would be significantly  shorter at higher redshift. On the other hand, this argument in absence of more complex factors should lead to a steeper, monotonic increase of \mhp that we do not observe.

As mentioned above, AGN feedback at high redshifts can also regulate galaxy star formation and explain our observed \mhp trend.
AGN activity at high redshift is expected to be almost exclusively in quasar mode \citep[e.g.,][]{SilkRees1998} with powerful outflows that can heat or even expel gas. 
However such radiative feedback has shown to be inefficient in  hydrodynamical ``zoom--in" simulations at $z\sim6$  \citep[e.g.][]{Costa2014}.  Observations also indicate that high-$z$ quasars do not prevent significant reservoirs of cold gas from  fuelling  star formation \citep[e.g.,][]{Maiolino2012,Cicone2014}. 
Therefore, star formation in massive haloes can proceed for $2\!-\!3$\,Gyr after the Big Bang without being  significantly affected by AGN activity, in agreement with our observations. At later times, perturbations to cold filamentary accretion can starve galaxies of their gas supplies \citep{Dubois2013a}.   

A deeper understanding of the role played by AGN comes from studying their co-evolution with  super-massive black hole (BH). \citet{Beckmann+2017} show that once re-normalised for the ratio between the BH mass ($M_\mathrm{BH}$) and the virial mass of the halo, the impact of AGN feedback is the same from $z=0$ to 5. According to their hydrodynamical simulations \citep[from the \textsc{Horizon-AGN} suite,][]{Dubois+14} this process is able to suppress  galaxy stellar mass assembly when $M_\mathrm{BH}/M_\mathrm{h}>4\times10^{-5}$.  In first approximation, this critical threshold is in good agreement with the one that can be derived from COSMOS if the critical BH mass ($M_\mathrm{BH,crit}$) is correlated with \mhpp. Assuming a BH-to-stellar mass ratio of $2\times10^{-3}$ \citep{MarconiHunt2003} we can write    
\begin{equation}
\frac{M_\mathrm{BH,crit}}{M_\mathrm{h}^\mathrm{peak}} =  \frac{M_\mathrm{BH}}{M_{\ast}} \times \left(\frac{M_\ast}{M_\mathrm{h}}\right)^\mathrm{peak} = 2\times10^{-3} \times 10^{-1.7 \pm 0.1},
\label{eq:mbh_crit}
\end{equation}
which gives $3\!-\!5\times10^{-5}$ including the variation in the \msmh ratio calculated at $M_\mathrm{h} \equiv M_\mathrm{h}^\mathrm{peak}$ (see Fig.~\ref{fig:SHMRCosmos_upto6}, \ref{fig:SHMRCosmos_789}, and \ref{fig:MsMh_fixMh}).  
At least at $z<2$ the anti-hierarchical  growth of the BH mass function  \citep[][]{Marconi+2004,Shankar+2009,Shankar+2013} implies that more massive BHs form earlier, so \mhp must also  increase  (as we find in COSMOS) in order to keep the ratio constant. 
In other words, if we assume that the  quenching threshold $M_\mathrm{BH,crit}/M_\mathrm{h}^\mathrm{peak}$ is universal, BH formation models can use the COSMOS  stellar-to-halo mass relationship as an indirect constraint. 
Part of the redshift evolution may  also be due to the ratio between $M_\mathrm{BH}$ and $M_\ast$. In Eq.~\ref{eq:mbh_crit} we used a constant value but other studies indicate that such a relationship varies depending on the galaxy bulge component \citep{ReinesVolonteri2015}. However, in the \textsc{Horizon-AGN} simulation this quantity has been shown to remain constant ($\sim\!2\times10^{-3}$)  up to $z=3$ \citep[][]{Volonteri+2016b}. 

Recently \citet{Glazebrook+2017} reported a  massive ($M_* = 1.7\times10^{11} M_\odot$) and quiescent galaxy at a spectroscopic redshift of $z=3.717$. This observation suggests a scenario where in the early Universe dark matter haloes are hosting massive star-forming galaxies and that the quenching of star formation appears as early as $z\sim4$. According to our SHMR relation this stellar mass corresponds to a halo mass of $M_{\rm h} = \sim 10^{12.5} M_\odot$ so around our value of \mhp for $z=4$. This observation is in agreement with our argument that \mhp is the characteristic mass for haloes currently undergoing quenching.

%Models of star formation efficiency discussed so far do not account for the role of large scale structures. 
Depending on their location within the cosmic web (filaments, nodes, voids) haloes with similar masses may experience  different accretion histories \citep{DeLucia+2012}.   
One key idea in this context is  ``cosmic web detachment'' \citep{Aragon2016}: Galaxies tied to nodes or filaments are removed from their original location by  interaction with another galaxy.   After the detachment gas supply -- and then star formation -- becomes less efficient. \citet{Aragon2016} suggest that massive haloes are the first to detach, whereas less massive haloes $0.1\!-\!3\times10^{10}\,h^{-1}\,M_\odot$ are still part of the cosmic web   today. It is difficult to test this scenario beyond the local universe  because precise measurements of the SMF are required in addition to higher-order statistics (e.g., 3-point correlation functions). We emphasize that COSMOS is the ideal laboratory to test the impact of large-scale environment in the models mentioned above, because the cosmic structure of this field has been reconstructed  at least up to $z\sim1$ \citep{Darvish+2014,Laigle+2018}. We aim to perform such an analysis in a future work.  

In summary, we have described different physical processes which could explain our observed trends of \mhp and the SHMR with redshift. Of course, in the real universe the truth is likely to be some combination of these mechanisms. But based on this discussion the physical processes at work in results seem to be best understood as a combination of cold-flow accretion and AGN feedback combined with anti-heirarchical growth of the black hole mass function, with the precise role of evolutionary and environmental effects yet to be determined. 

\section{Conclusions}

We have used a sub-halo abundance matching technique combined with precise stellar mass function measurements in  COSMOS to make a new measurement of the stellar-to-halo mass relationship from $z\sim0$ to $z\sim5$. We accounted for the principal sources of uncertainties in our stellar mass measurements and photometric redshifts. We also tested the impact of halo mass function uncertainties on the resulting SHMR. At $z\sim0.2$ we found that the ratio of mass in stars to dark matter halo mass (\msmhp) peaks at a halo mass of $10^{12.05 \pm 0.07} M_\odot$. This peak mass increases steadily to $10^{12.48 \pm 0.08} M_\odot$ at $z\sim2.3$, and remains almost constant up to $z=4$.

By comparing our results to studies that rely on models accounting for both central and satellite galaxies, we have shown that at least at $z<2$ the distinction between central and satellite galaxies has only a limited impact on the peak halo mass \mhpp. A complete modelling including satellite galaxies is left to a future work.

We found that the \msmh ratio has little dependence on redshift for haloes more massive than \mhpp, but strongly depends on redshift for less-massive haloes, consistent with the picture that the star formation has been quenched in massive haloes and continues in less massive haloes. We showed that the evolution of the shape of the stellar mass function has a strong impact on the the SHMR: the change in the position of the knee of the SMF is responsible for the shift in the value of \mhpp. Accurate SMF estimations at high redshift for massive galaxies are needed to constrain the SHMR. We also show how mass function uncertainties can influence our measurements of \mhpp. 

We discussed qualitatively which physical processes control the SHMR  and \mhpp, which we interpret as the characteristic mass of quenched haloes. We speculate that this evolutions can either be related to AGN feedback or to environmental effects such as cold gas inflows at high redshift and cosmic web detachment.

Our study is based on a phenomenological model and as such can provide no direct information concerning the physical processes acting inside haloes. Next-generation hydrodynamical simulations will allow us to better understand the small-scale physical processes acting inside dark matter haloes and determine what physical effects control star formation. In the next few years, the combined 20 ${\rm deg}^2$ Spitzer--Euclid legacy and Hawaii-2-0 surveys on the Euclid deep fields will provide much better constraints on the massive end of the SMF at high redshifts. Precise photometric redshifts will allow us to investigate in detail the role of environment and in particular the ``cosmic web" role in shaping galaxy and dark matter evolution. 

For future surveys like Euclid, errors on the cosmological figure-of-merit will be dominated by systematic errors. For this reason it is essential to understand the interplay between baryons and dark matter on small scales and the uncertainties present in estimates of the halo mass function.

\section*{Acknowledgements}
The COSMOS team in France acknowledges support from the Centre National d'\'{E}tudes Spatiales.
LL acknowledges financial support from Euclid Consortium and CNES.
HJM acknowledges financial support form the ``Programme national
cosmologie et galaxies'' (PNCG) and the DIM-ACAV. 
OI acknowledge funding of the French Agence Nationale de la Recherche for the ``SAGACE'' project.
Support for this work was also provided by NASA.
The Cosmic Dawn Center is funded by the DNRF. This research is also partly supported by the Centre National d'Etudes Spatiales (CNES).
The authors wish to thank P.\ Behroozi, J.\ Silk, M. Lehnert, G.\ Mamon, N.\ Malavasi, R.~M.~J.\ Janssen and G.\ Despali for their useful comments on an earlier version of this manuscript. The authors would also like to thank the referee for a thorough and constructive report which improved this paper.
This work is based on data products from observations made with ESO Telescopes at the La Silla Paranal Observatory under ESO programme ID 179.A-2005 and on data products produced by TERAPIX and the Cambridge Astronomy Survey Unit on behalf of the UltraVISTA consortium. 
It is also based on observations and archival data made with the Spitzer Space Telescope, which is operated by the Jet Propulsion Laboratory, California Institute of Technology under a contract with NASA.

%%%%%%%%%%%%%%%%%%%%%%%%%%%%%%%%%%%%%%%%%%%%%%%%%%



\bibliographystyle{mnras}
\bibliography{biblio} % if your bibtex file is called example.bib

\appendix

\section{Fitting the Bolsho\"{i}-Planck HMF for $M_{\rm max}$}

\begin{figure}
    \centering
    \includegraphics[width=0.99\columnwidth]{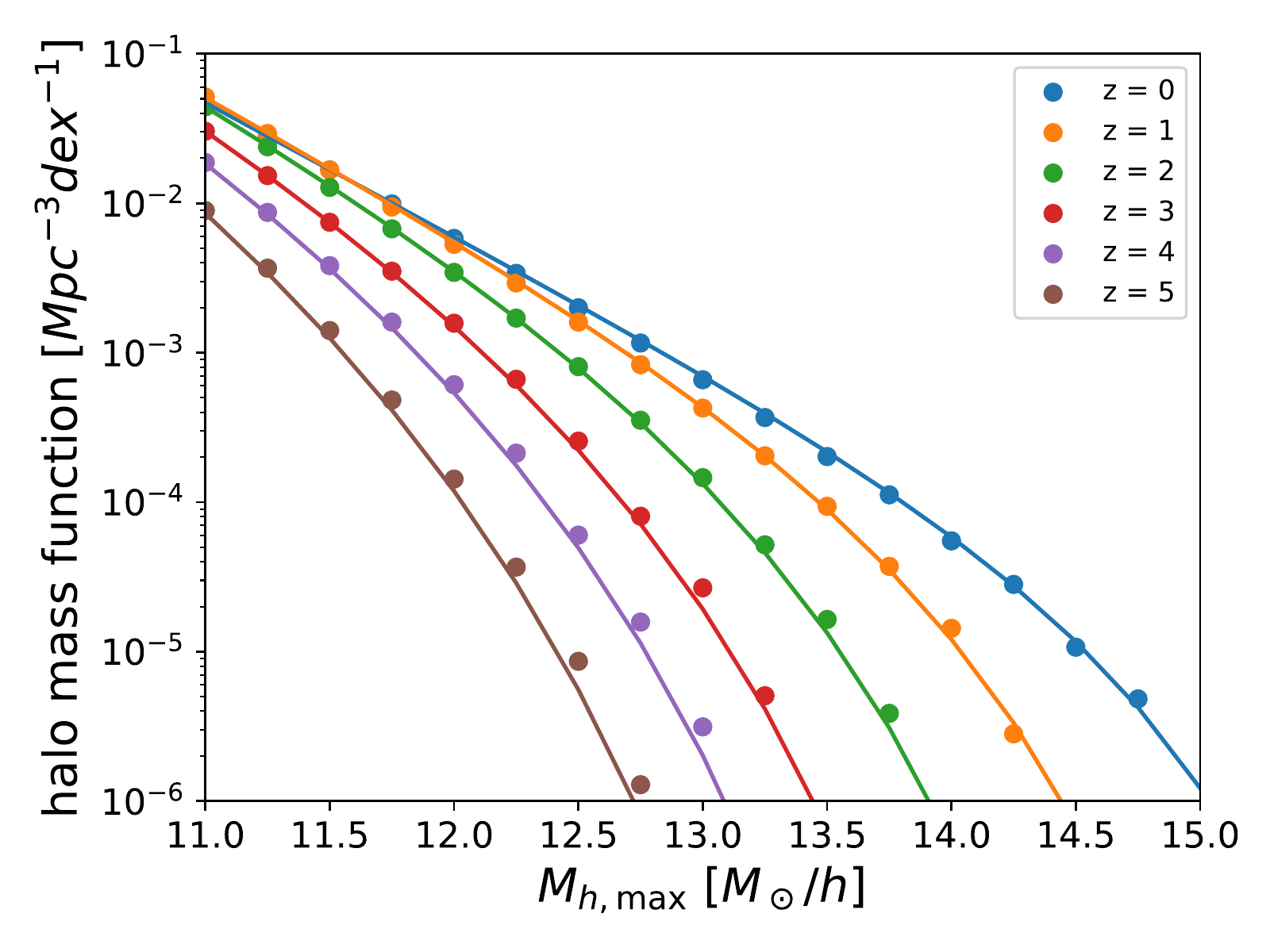}
    \caption{Points show halo densities obtained from the Bolsho\"{i}-Planck simulation for different redshift snapshots. The fit of the \citet{Despali+2016} HMF on this data points is shown as the plain lines.}
    \label{fig:despaliFit}
\end{figure}

Fig \ref{fig:despaliFit} shows the fit of the \citet{Despali+2016} HMF on the halo number densities of the Bolsho\"{i}-Planck simulation, using the maximal mass in the history of the haloes. See section \ref{sec:HMF}.

\section{MCMC best fit parameters}

\begin{figure*}
\includegraphics[width=0.49\textwidth]{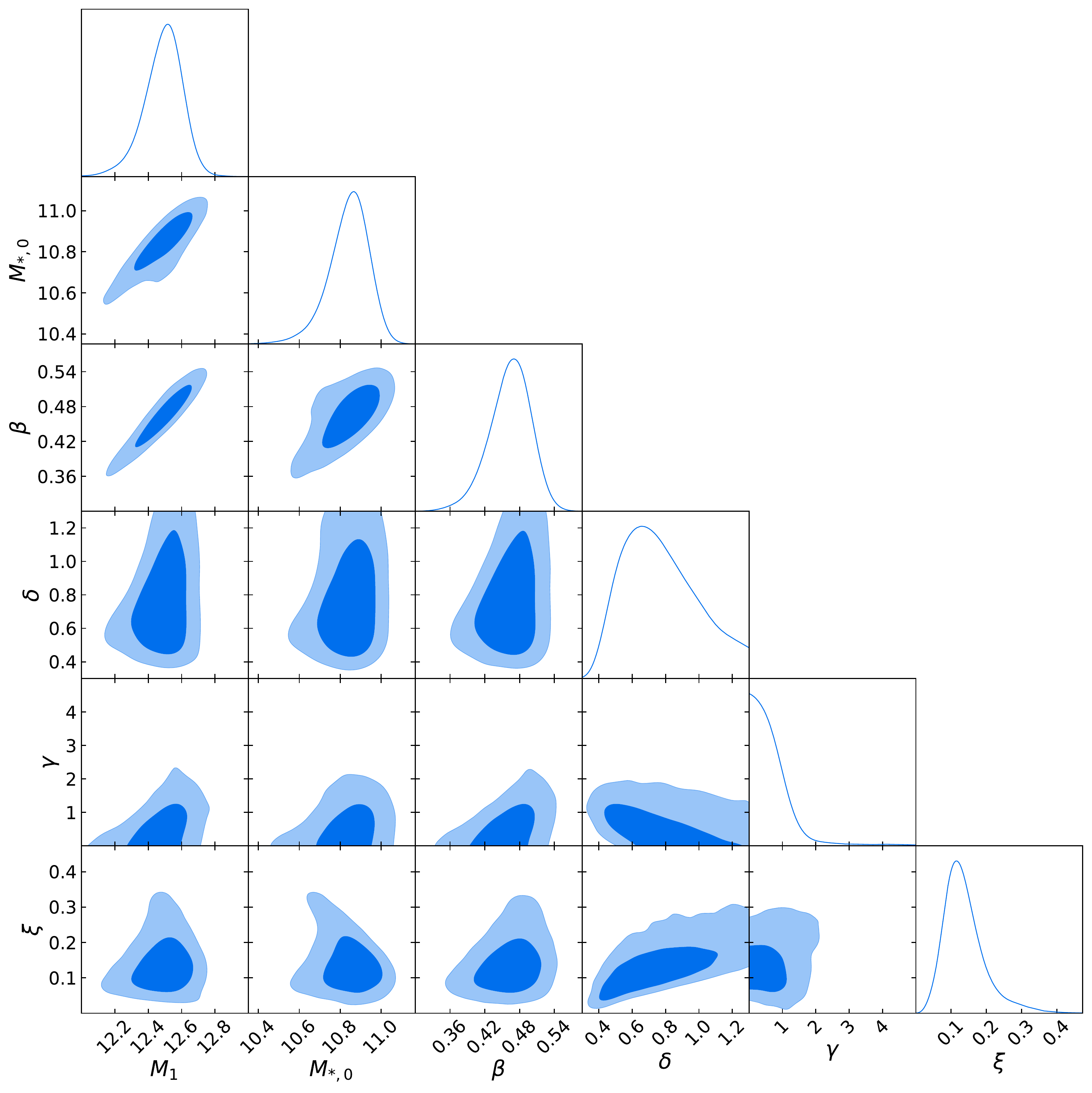}
\includegraphics[width=0.49\textwidth]{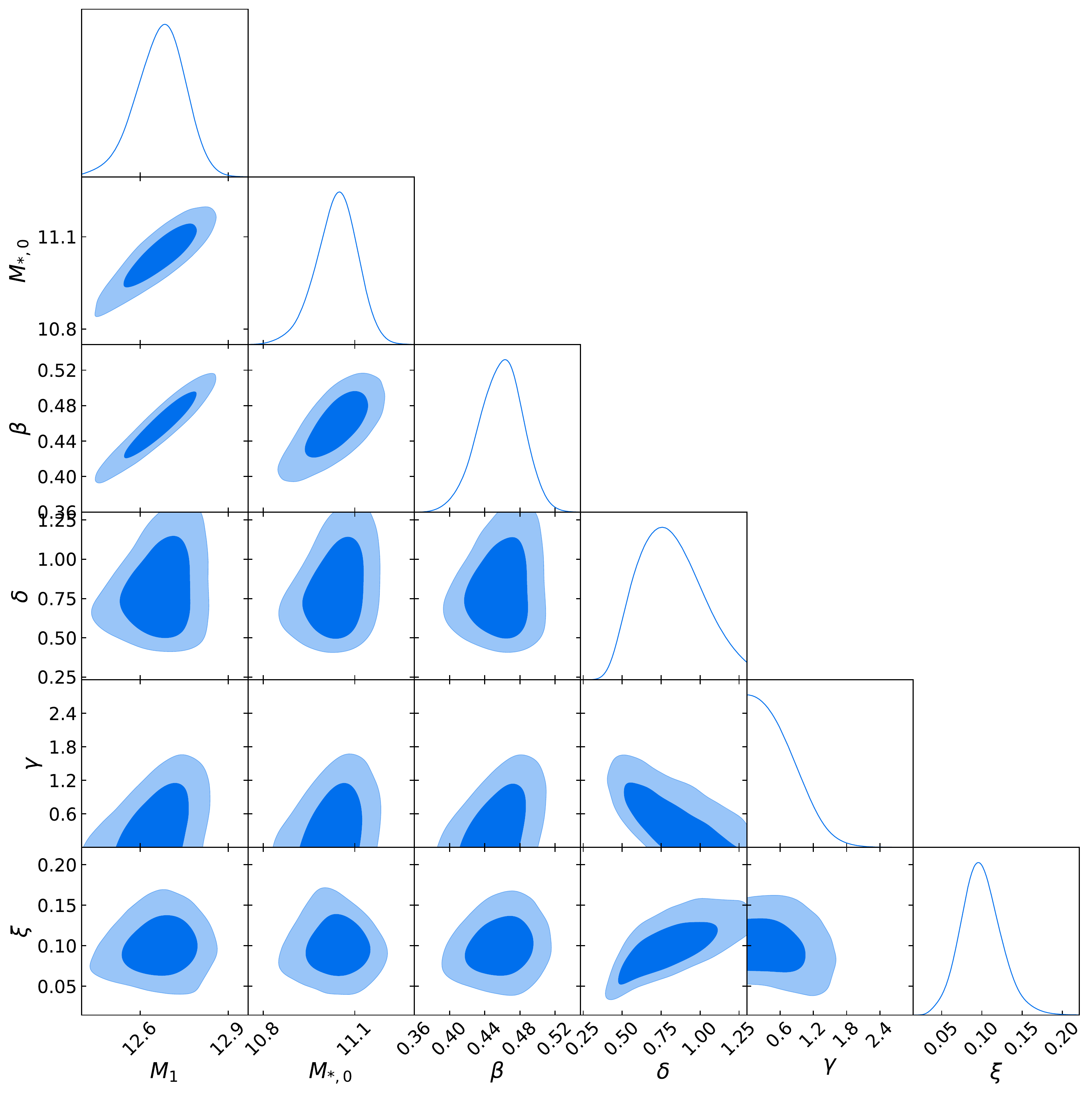}\\
\caption{ One and two dimensional marginalised distributions for the six free parameters. Solid contours give the 68 and 95 per cent confidence intervals. Left pannel is for redshift bin [0.2, 0.5], right pannel is for redshift bin [0.5, 0.8].}
\label{fig:post_0-1}
\end{figure*}

\begin{figure*}
\includegraphics[width=0.49\textwidth]{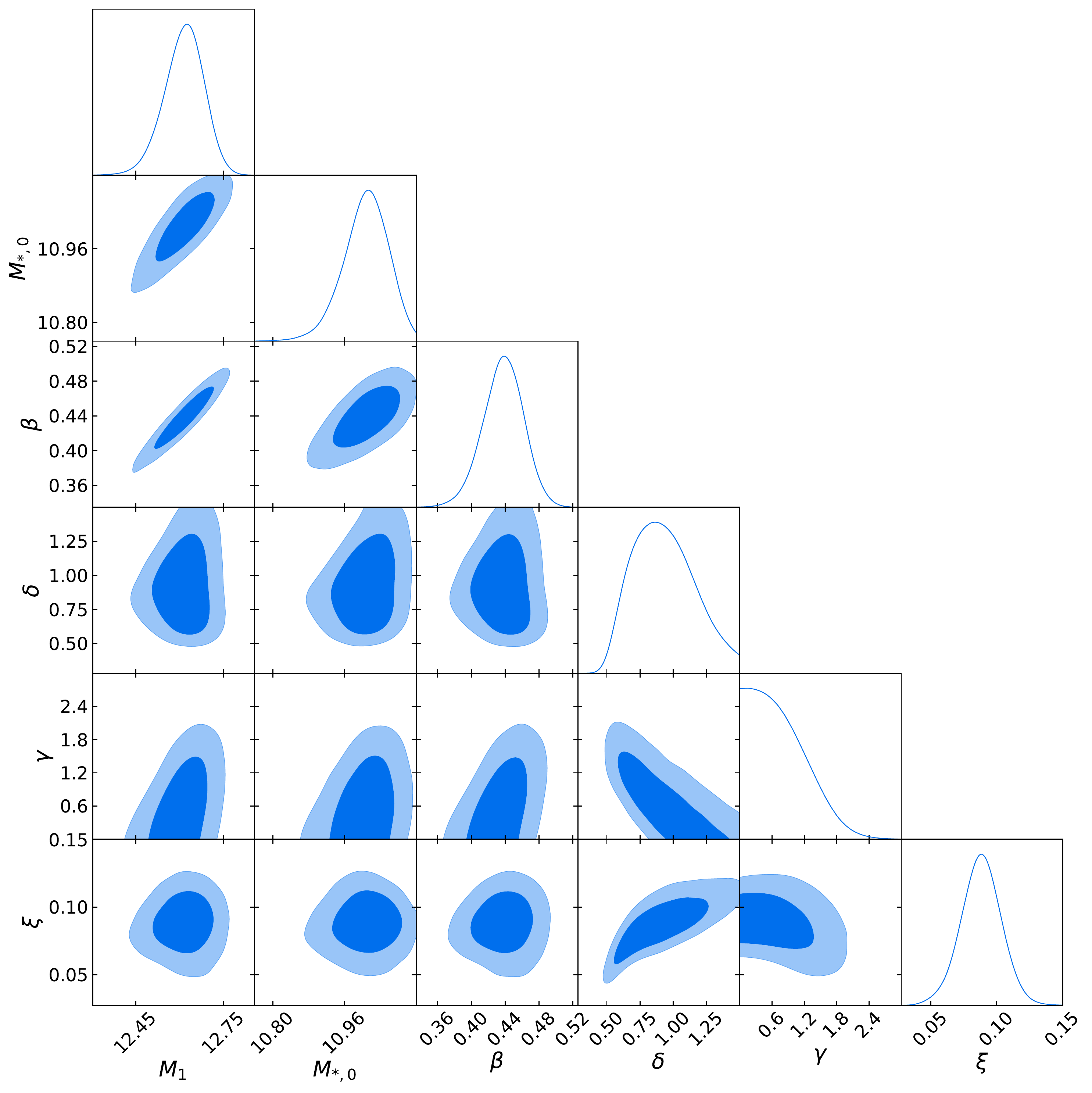}\includegraphics[width=0.49\textwidth]{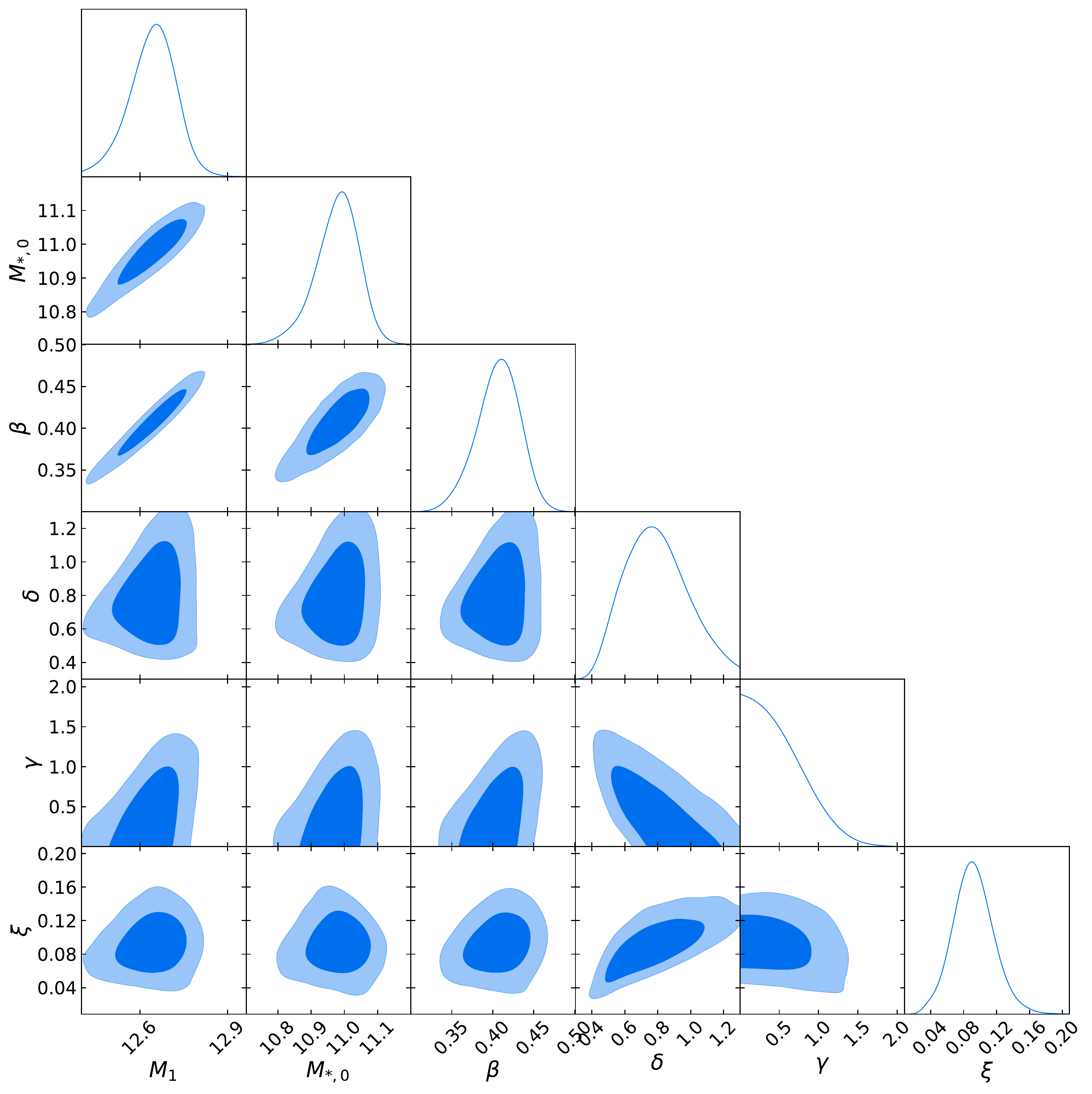}\\
\includegraphics[width=0.49\textwidth]{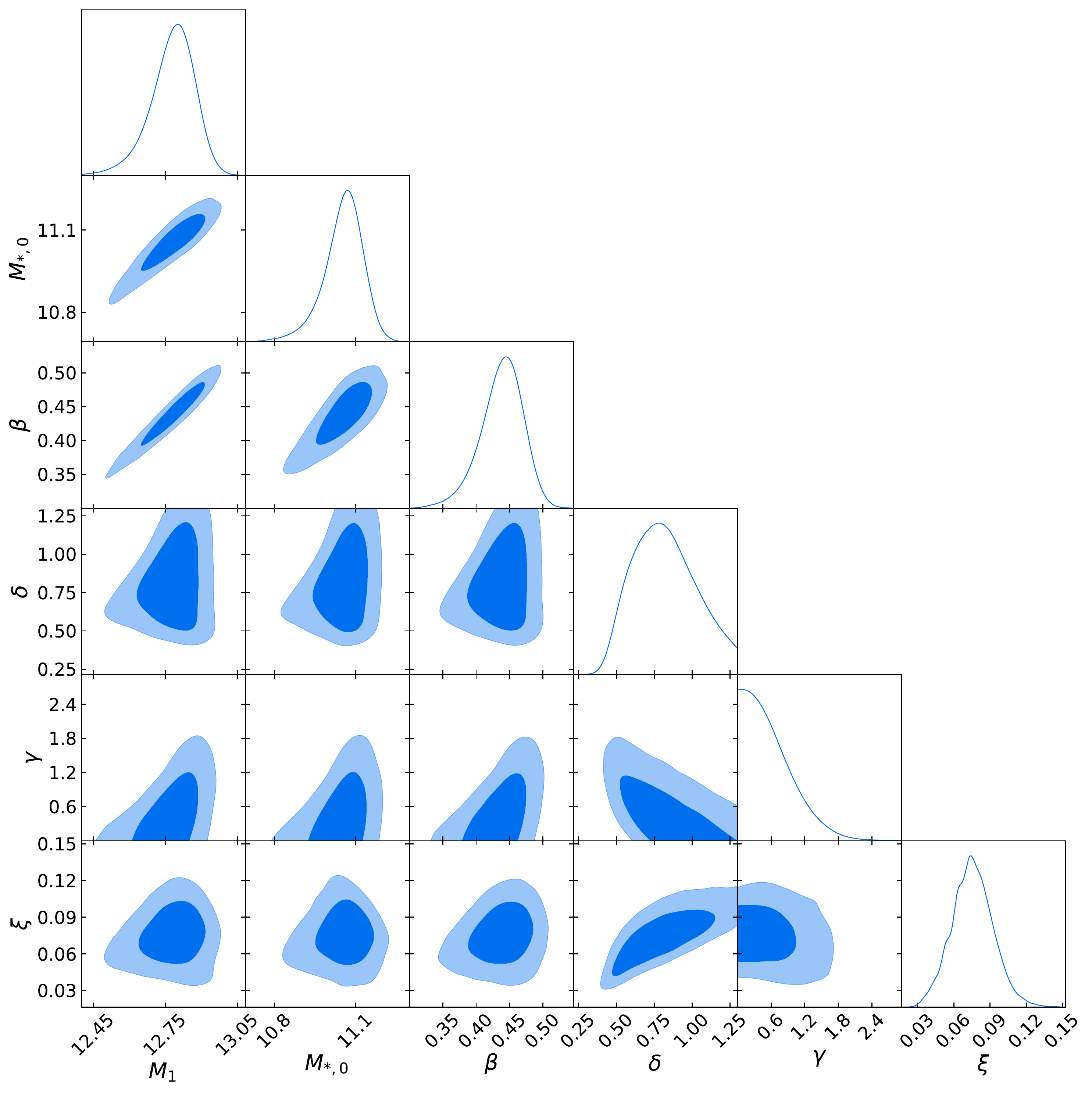}\includegraphics[width=0.49\textwidth]{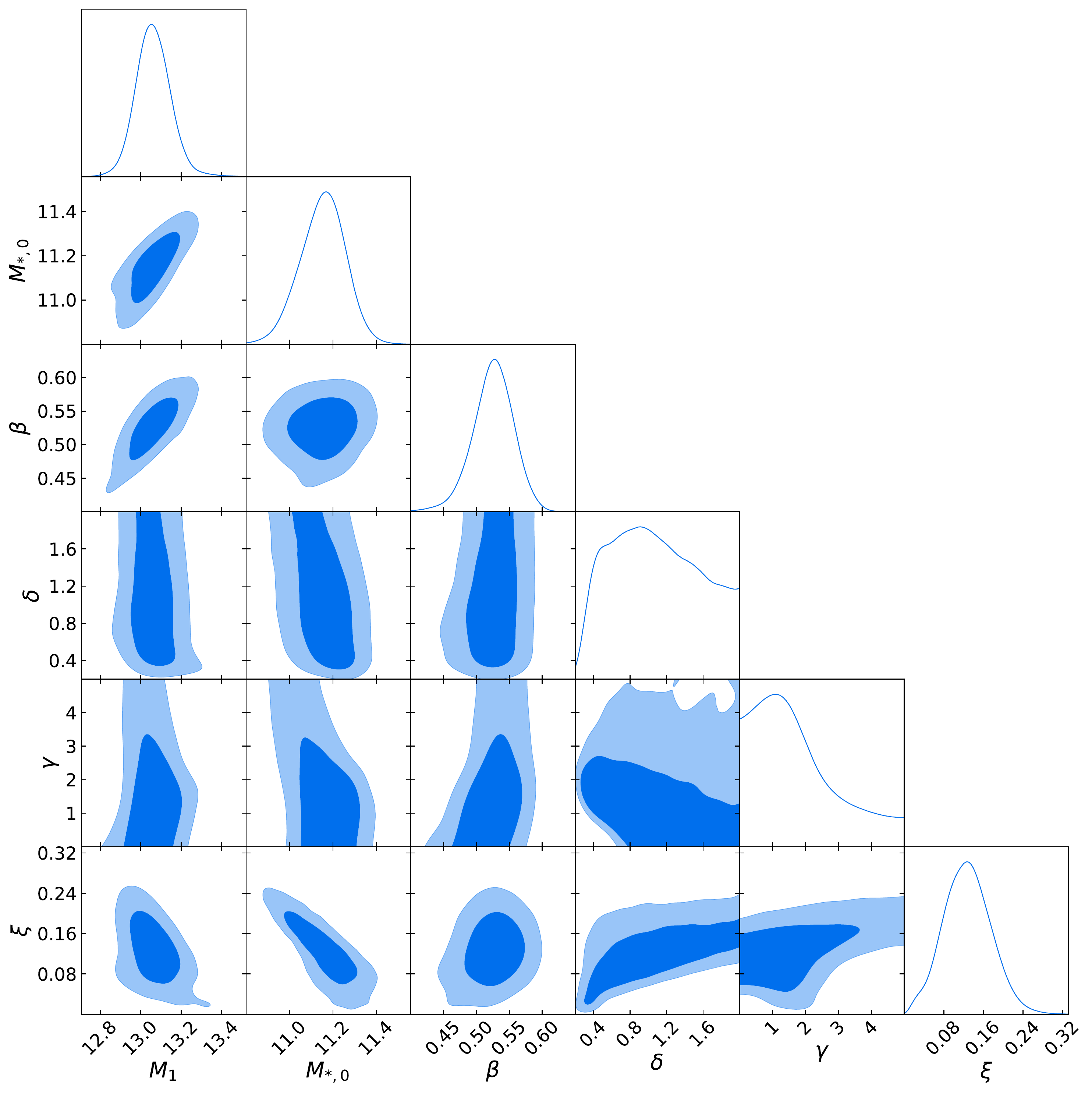}
\caption{Same as \ref{fig:post_0-1} for (from left to right and top to bottom) [0.8, 1.1], [1.1, 1.5], [1.5, 2], [2, 2.5] redshift bins.}
\label{fig:post_2-5}
\end{figure*}

\begin{figure*}
\includegraphics[width=0.49\textwidth]{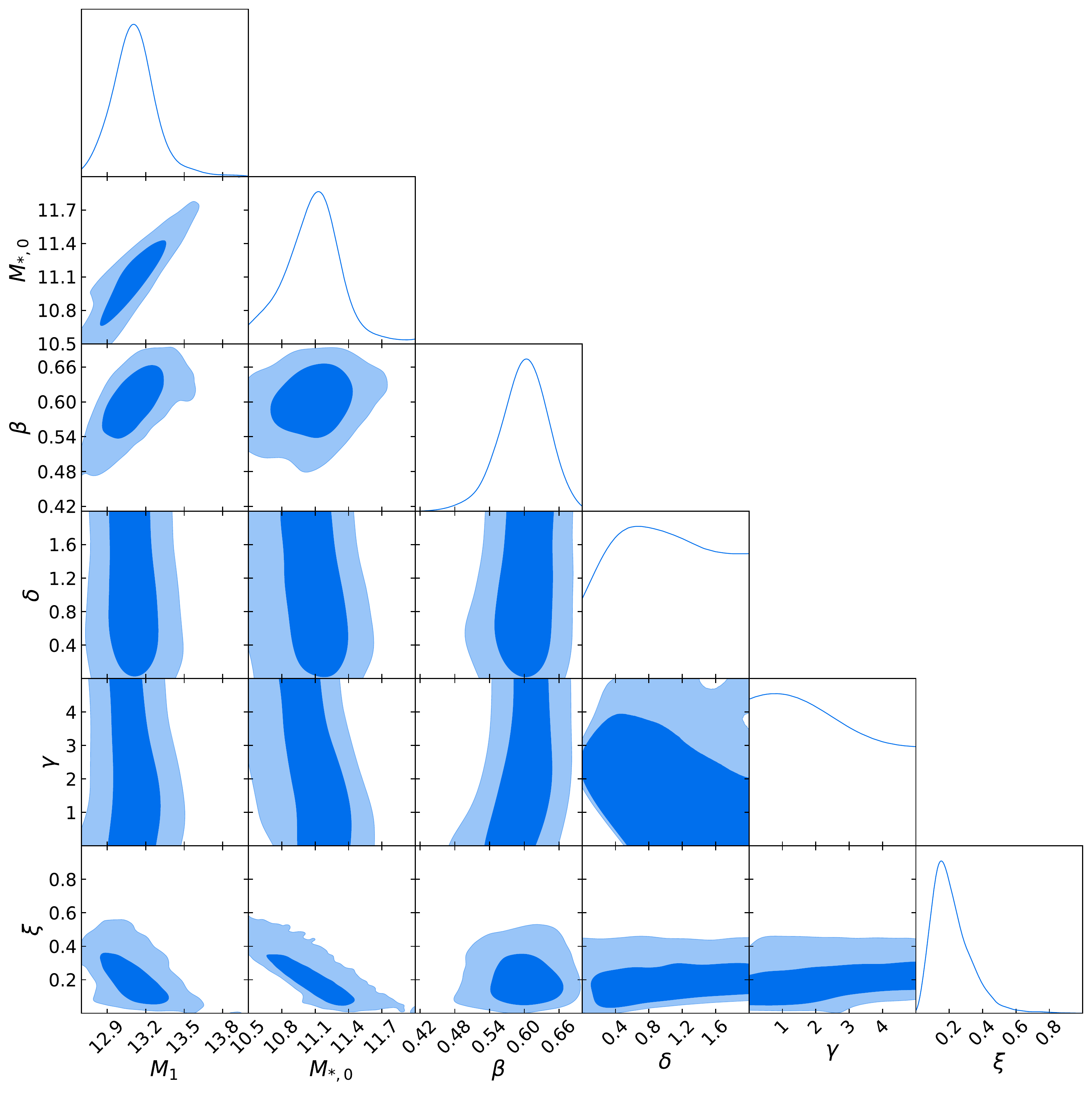}\includegraphics[width=0.49\textwidth]{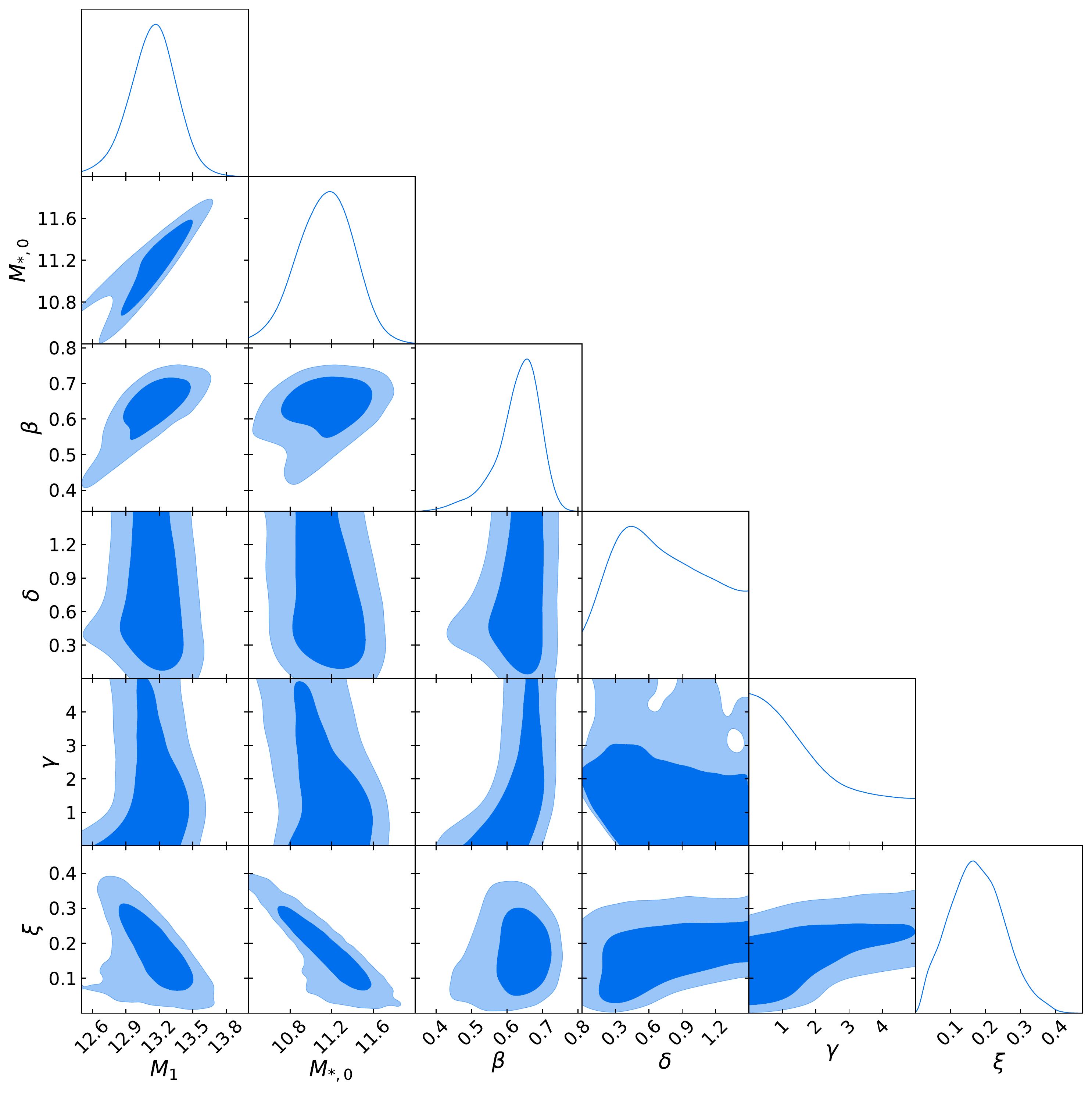}\\
\includegraphics[width=0.49\textwidth]{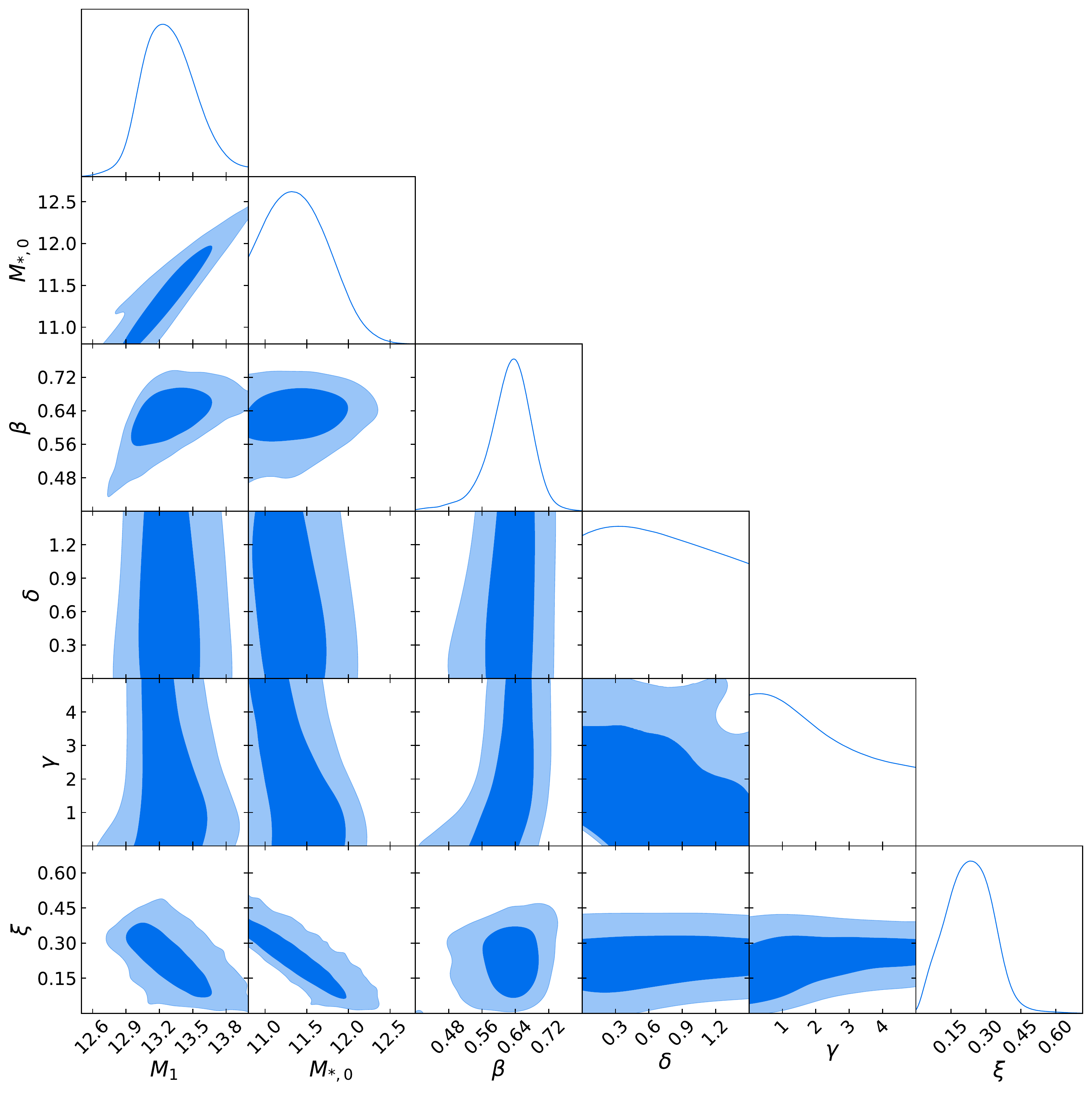}\includegraphics[width=0.49\textwidth]{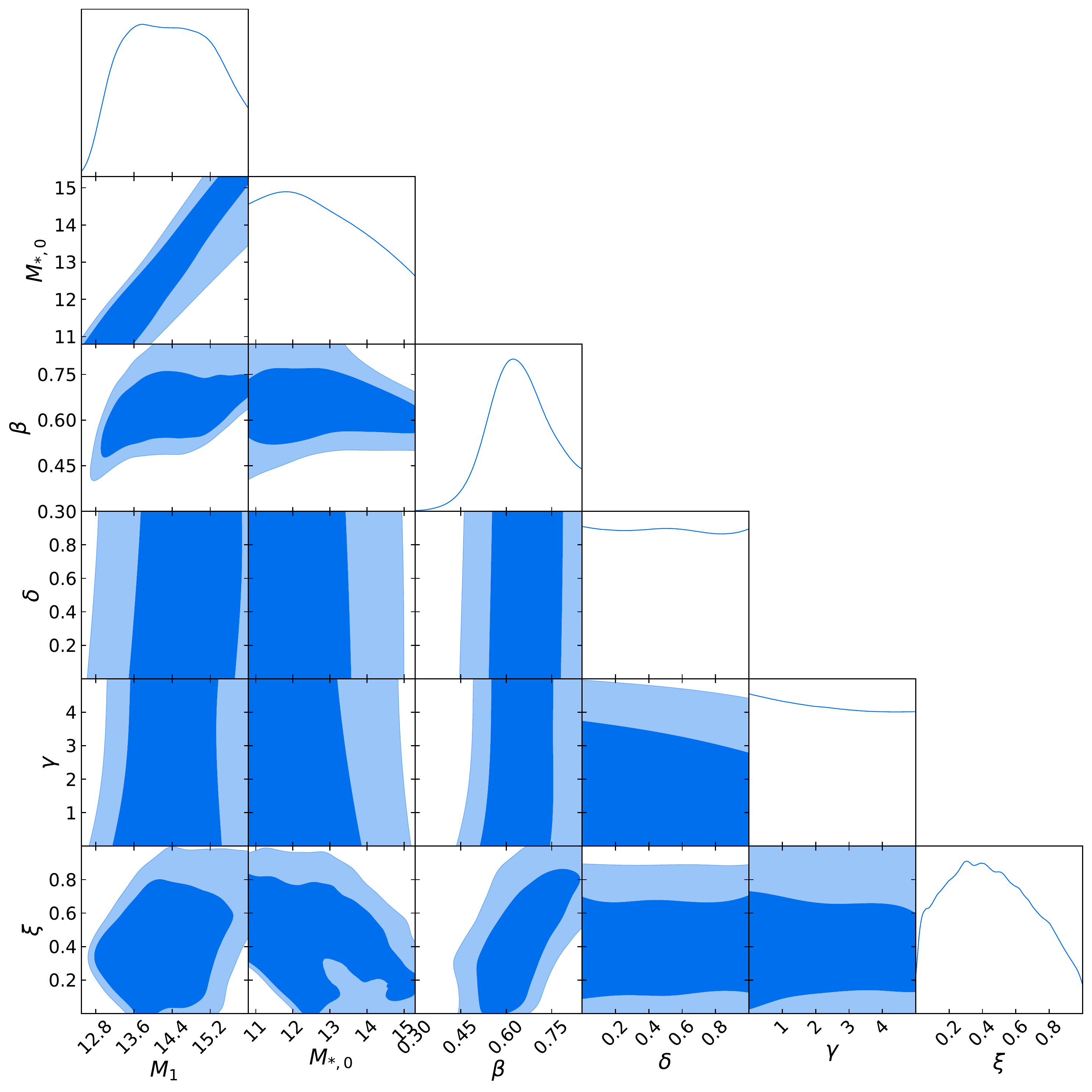}
\caption{Same as \ref{fig:post_0-1} for (from left to right and top to bottom) [2.5, 3], [3, 3.5], [3.5, 4.5], [4.5, 5.5] redshift bins.}
\label{fig:post_6-9}
\end{figure*}

In this Appendix we provide further details about the COSMOS2015 galaxy SMF (median $z$ and limiting stellar mass of each bin, see Table \ref{tab:survey_details}) and the best-fit parameters of Eq.~(\ref{eq:SHMR}) resulting from our MCMC method (see Table \ref{tab:MCMC_fits}). In addition, Fig. ~\ref{fig:post_0-1}-\ref{fig:post_6-9} show the MCMC  posterior distributions in the ten redshift bins independently considered in this analysis.

\begin{table}
\caption{Median redshift of each redshift bin and limiting stellar mass of COSMOS survey as defined in D17.}
\label{tab:survey_details}
\begin{tabular} {ccc}
\hline
\hline
Redshift bin & median $z$ & $\log(M_\mathrm{\ast,lim}/M_\odot)$ \\

\hline
(0.2, 0.5] &  0.370 & 8.17 \\

(0.5, 0.8] &  0.668 & 8.40 \\

(0.8, 1.1] &  0.938 & 8.58 \\

(1.1, 1.5] &  1.29 & 8.77  \\

(1.5, 2.0] &   1.74 & 8.98 \\

(2.0, 2.5] &  2.22 & 9.17 \\

(2.5, 3.0] &  2.68 & 9.32 \\

(3.0, 3.5] &  3.27 & 9.50 \\

(3.5, 4.5] &  3.93 & 9.67 \\

(4.5, 5.5] &  4.80 & 9.86 \\
\hline
\end{tabular}
\end{table}

{ % begin box to localize effect of arraystretch change
\renewcommand{\arraystretch}{1.5}
\begin{table*}
\caption{Best fit parameters for the ten redshift bins with their 68 per cent confidence intervals, and \mhp recovered from the best fit SHMR with its 68 per cent confidence interval.}\label{tab:MCMC_fits}

{\begin{tabular} {l l l l l l l l}
\hline
\hline

Redshift bin & $ \log(M_{1}/M_{\odot})$ & $ \log(M_{*,0}/ M_{\odot}) $ & $\beta$ & $\delta$ & $\gamma$ & $\xi$ & $ \log(M_{\rm h}^{\rm peak} /M_{\odot})$ \\

\hline
[0.2, 0.5] & $12.49^{+0.13}_{-0.094}$ & $10.84^{+0.11}_{-0.077}$ & $0.463^{+0.040}_{-0.030}$ & $0.77^{+0.16}_{-0.29}$ & $< 0.802$ & $0.138^{+0.034}_{-0.066}$ & $12.05 \pm 0.07$\\

[0.5, 0.8] & $12.668^{+0.089}_{-0.074}$ & $11.039^{+0.074}_{-0.060}$ & $0.458^{+0.026}_{-0.023}$ & $0.81^{+0.17}_{-0.24}$ & $< 0.723$ & $0.099^{+0.022}_{-0.027}$ & $12.22 \pm 0.05$\\

[0.8, 1.1] & $12.614^{+0.073}_{-0.060}$ & $11.006^{+0.056}_{-0.042}$ & $0.437^{+0.025}_{-0.022}$ & $0.93^{+0.19}_{-0.28}$ & $< 0.955$ & $0.088\pm 0.015$ & $12.14 \pm 0.04$\\

[1.1, 1.5] & $12.642^{+0.086}_{-0.069}$ & $10.978^{+0.072}_{-0.054}$ & $0.407^{+0.029}_{-0.023}$ & $0.80^{+0.16}_{-0.23}$ & $< 0.629$ & $0.092^{+0.023}_{-0.025}$ & $12.26 \pm 0.04$\\

[1.5, 2.0] & $12.78^{+0.10}_{-0.072}$ & $11.053^{+0.080}_{-0.055}$ & $0.438^{+0.035}_{-0.026}$ & $0.82^{+0.17}_{-0.25}$ & $< 0.724$ & $0.075\pm 0.017$ & $12.35 \pm 0.04$\\

[2.0, 2.5] & $13.062^{+0.078}_{-0.087}$ & $11.15^{+0.11}_{-0.095}$ & $0.525^{+0.033}_{-0.027}$ & $1.09^{+0.36}_{-0.68}$ & $< 2.08$ & $0.128^{+0.045}_{-0.050}$ & $12.48 \pm 0.08$\\

[2.5, 3.0] & $13.11\pm 0.18$ & $11.09\pm 0.25$ & $0.598^{+0.045}_{-0.036}$ & $1.01^{+0.55}_{-0.72}$ & $---$ & $0.216^{+0.061}_{-0.14}$ & $12.47 \pm 0.19$\\

[3.0, 3.5] & $13.14^{+0.22}_{-0.20}$ & $11.14\pm 0.27$ & $0.631^{+0.071}_{-0.038}$ & $0.73^{+0.35}_{-0.54}$ & $< 2.47$ & $0.176^{+0.074}_{-0.085}$ & $12.49 \pm 0.17$\\

[3.5, 4.5] & $13.30^{+0.20}_{-0.27}$ & $11.41^{+0.28}_{-0.46}$ & $0.625^{+0.056}_{-0.039}$ & $---$ & $< 2.93$ & $0.231\pm 0.099$ & $12.63 \pm 0.25$\\

[4.5, 5.5] & $14.35^{+0.89}_{-1.0}$ & $< 13.5$ & $0.642^{+0.094}_{-0.11}$ & $---$ & $---$ & $0.45^{+0.22}_{-0.34}$ & $13.35 \pm 0.54$\\

\\

\hline
\end{tabular}}
%\tablefoot{SHMR parameters estimation and confidence intervals  from MCMC sampling.}

\end{table*}
} % end arraystretch box

\newpage

% Don't change these lines
\bsp	% typesetting comment
\label{lastpage}
\end{document}